\documentclass[12pt,preprint]{aastex}

\newcommand{\eg}{e.\,g.\,}

\input epsf
\usepackage{rotating}
\usepackage{float}
\shorttitle{LCOGT}
\shortauthors{Brown}

\begin{document}
\title{Las Cumbres Observatory Global Telescope Network}
\author{Brown, T.M.,
Baliber, N.\altaffilmark{1}, 
Bianco, F.B.\altaffilmark{2}, 
Bowman, M., 
Burleson, B., 
Conway, P., 
Crellin, M., 
Depagne, \'E.\altaffilmark{3}, 
De Vera, J., 
Dilday, B.,
Dragomir, D., 
Dubberley, M.\altaffilmark{4}, 
Eastman, J.D., 
Elphick, M., 
Falarski, M., 
Foale, S., 
Ford, M., 
Fulton, B.J.\altaffilmark{5}, 
Garza, J., 
Gomez, E.L., 
Graham, M.,
Greene, R., 
Haldeman, B., 
Hawkins, E., 
Haworth, B., 
Haynes, R.,
Hidas, M.,
Hjelstrom, A.E., 
Howell, D.A., 
Hygelund, J., 
Lister, T.A., 
Lobdill, R., 
Martinez, J., 
Mullins, D.S., 
Norbury, M., 
Parrent, J., 
Paulson, R., 
Petry, D.L., 
Pickles, A., 
Posner, V., 
Rosing, W.E., 
Ross, R., 
Sand, D.J.\altaffilmark{6}, 
Saunders, E.S., 
Shobbrook, J., 
Shporer, A.\altaffilmark{7}, 
Street, R.A., 
Thomas, D., 
Tsapras, Y., 
Tufts, J.R., 
Valenti, S., 
Vander Horst, K., 
Walker, Z., 
White, G., 
Willis, M.} 

\affil{Las Cumbres Observatory Global Telescope Network, 6740 Cortona Dr. Suite 102, Goleta, CA 93117, USA}
\altaffiltext{1}{Department of Astronomy, California Institute of Technology, Pasadena, CA 91125-0001}
\altaffiltext{2}{Department of Physics, New York University, N.Y, N.Y., 10012}
\altaffiltext{3}{Leibniz-Institut f\"ur Astrophysik, Potsdam, Germany}
\altaffiltext{4}{Santa Barbara Infrared, Inc., Santa Barbara, CA}
\altaffiltext{5}{Institute for Astronomy, University of Hawaii, Honolulu, HI 96822}
\altaffiltext{6}{Physics Department, Texas Tech University, Lubbock, TX, 79409-1051}
\altaffiltext{7}{Division of Geological and Planetary Sciences, California Institute of Technology, Pasadena, CA 91125}
\begin{abstract}
Las Cumbres Observatory Global Telescope (LCOGT) is a young
organization dedicated to time-domain observations at optical and (potentially)
near-IR wavelengths.
To this end, LCOGT is constructing a world-wide network of telescopes,
including the two 2m Faulkes telescopes,
as many as 17$\times$1m telescopes, and as many as 23$\times$40cm telescopes.
These telescopes initially will be outfitted for 
imaging and (excepting the 40cm telescopes) spectroscopy at wavelengths
between the atmospheric UV cutoff and the roughly 1-$\mu$m limit of silicon
detectors.
Since the first of LCOGT's 1m telescopes are now being deployed,
we lay out here LCOGT's scientific goals and the requirements that these
goals place on network architecture and performance, we 
summarize the network's present and projected level of development,
and we describe our expected schedule for completing it.
In the bulk of the paper, we describe in detail
the technical approaches that we have adopted to attain the desired performance.
In particular, we discuss our choices for the number and location of network
sites, for the number and sizes of telescopes, for the specifications of the
first generation of instruments, for the software that will schedule and
control the network's telescopes and reduce and archive its data, and
for the structure of the scientific and educational programs for which
the network will provide observations.
\end{abstract}


\section{INTRODUCTION}\label{introduction}

For about the last century, the dominant line of development in astronomical
facilities has been to build larger and larger telescopes to observe
fainter and fainter objects.
This approach has been spectacularly successful, but the cost of large
boundary-pushing telescopes has grown to the point that obtaining funding
for them may soon become impractical.
Thus, it is desirable to look for observing strategies that permit 
less expensive ways to learn about the cosmos.
One such approach goes under the rubric of ``time-domain astronomy'';
the idea is to exploit the temporal variability of astronomical sources of
radiation to learn something about their physical structures and their
interactions with other objects.
Of course, time-domain astronomy is not new.
Indeed, in the sense of positional astronomy (e.g., planetary motions),
such studies go back to antiquity. 
But in the modern era, scheduling practices have not been conducive to
observing programs that require (say) weeks of continuous 
observation of a single
target, or near-instant response to triggers from short-lived transient 
phenomena, or even relatively sparse time coverage of every one of a large
list of targets.
With great effort, workers have arranged campaigns
to monitor short-period variability of stars over long time spans
(e.g., the Whole Earth Telescope, \citet{1990ApJ...361..309N},
or the SONG network, now in construction, for 
asteroseismology and microlensing
\citep{2008JPhCS.118a2041G}),
and by building special-purpose small telescopes, 
$e.g.$ the ETC, LOTIS, ROTSE, and RAPTOR systems, to observe gamma-ray bursts
\citep{1992ASPC...34..123V, 1998SPIE.3355..658P, 2000AJ....119.1901A, 
2002SPIE.4845..126V} and PAIRITEL \citep{2006ASPC..351..751B}, 
which in addition performs 
near-infrared observations of many other kinds of phenomena.
But such projects are hampered by the lack of 
flexible facilities that can
provide time-sampled data over a range of temporal cadences, dataset durations,
and observing modes, in a routine and systematic way.
Las Cumbres Observatory Global Telescope (LCOGT) is such a facility.
It is the first general-purpose, flexibly-scheduled, multi-instrument
optical observatory designed expressly to pursue astronomical
research in the time domain.
Its success depends not so much on advancing the state of the art in 
telescope technology
as on deploying a global network of telescopes that exploits all the 
communication, coordination, automation,
and data-processing strategies made possible by modern computing networks.
The combination of robotic telescopes with internet communication 
proves to be a potent one, and one that should bring major advances in some
branches of astronomical observing.

Here we describe LCOGT's aims, and the hardware
and software systems that we have built to meet them. 
The remainder of the paper is organized as follows.
Section 2 gives a very brief description of the scientific opportunities
offered by time-domain astronomy.
Section 3 describes the general structure -- number and geographic distribution
of sites, number and size of telescopes -- that we believe necessary 
for a network of ground-based telescopes to
exploit these opportunities.
In Section 4 we describe the telescopes that we have deployed, or will soon
deploy, to various nodes of the LCOGT network.
We place particular emphasis on our network of 1m telescopes,
because they represent the most novel part of the growing network, and because
they provide the largest portion of the network's observing capability.
Section 5 describes the design and capabilities of the instruments 
(imaging and spectrographic) that
are now installed on the network telescopes.
Section 6 is a brief description of the distributed software system that
schedules and assigns observing tasks to particular telescopes,
that performs initial reductions both for real-time quality assurance and
as input for later scientific interpretation, and that archives the result
for access by users of the LCOGT Network (henceforth, ``the Network'') 
and by the public.
We note that the Network's most novel and distinctive feature is the
software system that organizes its functions.
In this paper we describe this system at a high functional level;
much more detailed descriptions of its major components may be found
in \citet{2010SPIE.7737E..17H}, and in \citet{2010SPIE.7733E..90P, 2012SPIE.8444E..5VP}.
Section 7 describes our experience with shipping telescopes to their
intended sites and bringing them through the process of deployment,
start-up, and commissioning.
In Section 8 we outline the goals and some of the methods and accomplishments
of LCOGT's Education Program, which is built around ready access to the
observing resources of the Network.
Section 9 displays the first scientific result from our first
remote Network 1m telescope, obtained within weeks of its initial installation
at the McDonald Observatory in Texas, in April 2012, and
Section 10 gives a brief summary and describes prospects for the future. 

\section{ASTRONOMY IN THE TIME DOMAIN}

From millisecond pulsars to binaries with periods of hundreds of
years, astrophysical transients vary on timescales that span many
orders of magnitude.  Transient apparent brightnesses also range
from naked-eye to well beyond the detection limit of current
telescopes.  Even if restricted to optical light, no one facility
could be built to study all such phenomena.  In this section we detail
the characteristics of some of the most scientifically interesting
transients and variable sources, which we used to optimize the design of a global network
for studying time domain astronomy.

{\bf Dense time series:} Planetary transits of a star and stellar
variability have timescales of minutes to hours, requiring dense
time-series monitoring \citep{2000ApJ...529L..45C}.  High
signal-to-noise ratio (SNR) data are necessary to recover sub-1\% changes in
brightness. But because detection and follow-up methods for such objects
are biased towards bright stars, the bulk of known targets are fairly 
bright, often
brighter than 12th magnitude.  For these, moderate-aperture telescopes
provide the requisite photometric precision.  
Continuous observations are also important, to
prevent aliasing when periodograms are constructed.  This is
problematic at single-site facilities, where a given target can normally
be observed for 10 or fewer hours out of each 24.

{\bf Episodic time series:} Most supernovae rise and fall on a
timescale of months.  Many are powered by the radioactive decay of
$^{56}$Ni to $^{56}$Co (half-life 6.1 days) and $^{56}$Co to $^{56}$Fe
(half-life 77 days), so observations are generally required only every
few days for the reconstruction of a lightcurve.  The exception is at
early times, \eg\ catching the shock breakout from a red giant, which
happens in hours.  Just after explosion, the SN photosphere is also
expanding and doubling on a similar timescale.  
It is now rare to catch a SN explosion so early (e.g., SN2012fe in M101),
but with improved monitoring this will happen more frequently in the future.
In such cases, it is
necessary to observe a supernova as soon after discovery as is
feasible, which is greatly facilitated by the longitude distribution of 
a global network.  And because supernovae
are sometimes used as distance indicators, precise multi-band absolute
photometry is often required.  Since they are always located in
distant galaxies, the targets are rarely brighter than 14th magnitude.
Therefore, substantial observing time is required to build supernova
lightcurves, favoring a network composed of many nodes.

{\bf Triggered dense time series:} Microlensing of stars by planets
presents a unique challenge -- a huge number of stars along a densely
populated
line of sight must be monitored for signs of variability, and when an
anomalous microlens event is detected, dense time sampling with a cadence of
minutes must be activated.  A large number of telescopes distributed
in longitude would ensure that observations are possible in the narrow
time window when an anomaly is active.

{\bf Gamma-Ray Bursts:} The optical counterparts of gamma-ray bursts
(GRBs) fade as a power law on timescales from minutes to hours
\citep{2009ARA&A..47..567G}.  Departures from power-law behavior
(flares or breaks) are also observed, and carry information about such
explosion parameters as the jet opening angle.
Therefore, the most effective way to
study bursts is to have a network of automated telescopes with a rapid
response time and similar photometric response, 
well distributed in longitude.  Since
GRBs are at cosmological distances, telescope apertures larger than 1m
are preferable.

{\bf Very rapid time series:} Solar system bodies (e.g. Kuiper belt
objects) may take only fractions of a second to transit a star \citep[\eg
][]{2010Natur.465..897E}.  Characterizing their size requires multi-Hz
observations at multiple sites to probe multiple chords across the
occulting body.  

{\bf Lucky imaging:} Fast photometry cameras also enable ``lucky''
imaging, where many images per second are obtained, and only the $\sim
1\%$ least affected by seeing are used to resolve small angles
\citep{2006A&A...446..739L}.

{\bf Astrometry:} Near-Earth Objects (NEOs) are routinely discovered by large
surveys, but require targeted follow-up within days,
to provide astrometric data for accurate orbital elements before 
the object is lost to view.
Each of many targets require only small
amounts of time, making such work ideal for a robotic
facility.

{\bf Classification spectra of supernovae:} Thousands of supernovae
are discovered per year, but spectroscopic observations must be done
to determine redshifts and to classify them.  Since these extragalactic
sources can be faint, a high-efficiency spectrograph on a
large-aperture telescope is desired.  Only low resolution spectra are
necessary, thanks to rapid (thousands of km s$^{-1}$) expansion
velocities in SNe.  Since nearby supernovae usually have their peak output
in the optical, but have important lines distributed from the blue to
the red \citep{1997ARA&A..35..309F}, an ideal spectrograph would be
low resolution, have high efficiency, and cover the entire optical
range.
There is much to learn from time-resolved spectra of supernovae, but
even classification spectra benefit from the ability to respond quickly
to new discoveries.

{\bf RV vetting of transiting planet candidates: } The dimming of
starlight due to planetary transits can be mimicked by grazing binary
stars.  However, such transits also induce reflex motions in
their parent star as they orbit a common center of mass, which can be
detected by monitoring the star's radial velocity
\citep{1995Natur.378..355M}.  
The reflex signals of planets are at most a few hundred m s$^{-1}$,
while those of binary stars are typically many km s$^{-1}$.
Thus, accuracy better than about a hundred m s$^{-1}$ is necessary,
requiring a medium-to-high resolution spectrograph ($R \geq 40,000$).
The target stars are often bright, obviating the need for a large
aperture telescope, but multiple observations, distributed in time,
are needed to detect orbital motion.
In the near future, NASA's TESS mission \citep{2010AAS...21545006R} 
will provide many thousands
of candidate planets circling bright stars, all requiring such observations. 
Higher-precision radial velocities (a few m s$^{-1}$) allow one to determine
masses of transiting planets down to Neptune's mass or smaller.

{\bf The future:}  We doubt that all of the interesting applications of
time-domain observation have yet been explored, in part precisely because
there are, as yet, no suitable observing facilities to support them.
We hope that, going forward, the LCOGT network will inspire astronomers
to find applications that are not yet envisioned.

\subsection{Education and Public Engagement Activities}

From its inception, LCOGT has been committed to using astronomy 
as a tool for public engagement and informal education, 
with the aim of encouraging technically-minded learners of all ages 
to think critically and develop investigative skills. 
We have, however, found it surprisingly difficult to translate 
this commitment into effective action in the context of the public schools. 
The obstacle is that the resource that LCOGT can offer in 
relative abundance is telescope time, 
and not (as would be more traditional for a nonprofit education organization) 
support for teacher training, nor for curriculum development.

In consequence, we concentrate our educational efforts in two areas 
in which we can be effective. 
First, LCOGT provides access to observing time on its telescopes, 
to schools, to astronomy clubs and similar organizations, 
and sometimes to individuals. 
Some of this time is used in the form of real-time observing, 
in which users operate telescopes remotely. 
Recently, however, we have provided a larger portion as 
queue-scheduled observing, just as for science observations. 

Secondly, we have invested significant effort in 
the development of online tools and activities for education and 
public engagement. 
LCOGT develops programs for “citizen scientists”, 
people irrespective of their training or experience, who can become 
knowledgeable enough that they are able to assist in 
(and get recognition for) obtaining and analyzing observations that may be 
used to carry out genuine scientific investigations. 
These tools are necessarily web-based, hence they have the potential to 
reach people with a diverse range of abilities and backgrounds, 
and in large numbers.  
Some of these tools complement telescope use, whereas others provide a rich 
web-based experience for a general audience, as is discussed more in section 
8.

\section{NETWORK STRUCTURE --  OVERVIEW}

What features must a telescope network have, to meet the observing
challenges suggested above?
This question was prominent in our thinking from LCOGT's beginning;
to help us address it, we have relied heavily on discussions with the
community, especially as embodied in our advisory structure.
This structure consists of a Board of Trustees, an additional
small group of Counselors, who serve with no
definite term, and a much larger rotating-membership 
Scientific Advisory Committee.
With the help of these advisors, we defined both the broad science goals
to be pursued, and the key features that a network must have in order
to achieve the goals.
We address several of these features below, notably the number and geographical
distribution of nodes, the number and size of telescopes, the instruments
that are attached to those telescopes, and the software for coordinating
network observations and for making reduced data available.

Our description of these features is complicated by the
incompleteness of the LCOGT Network at the time of this writing,
and uncertainties regarding future funding, and therefore the
speed of construction and ultimate extent of the Network's facilities.
For clarity's sake, we thus describe the Network and its evolution in terms of 
3 completion Phases, which we label A, B, and C.
Phase A consists of all of the Network features that are operating as of 
the date of this writing (April 2013).
Phase B consists of additional facilities that LCOGT has funding to deploy,
and that we expect to complete before roughly January 2014.
The union of features in Phases A and B is thus a reasonable
expectation for the minimum capabilities of the completed Network.
These phases have been financed mostly by contributions from the TABASGO
Foundation (our primary donor), with generous additional funding from
the Scottish University Physics Alliance (SUPA), administered through
St. Andrews University.
Phase C consists of facilities beyond Phase B, 
which we would like to deploy as soon as possible, but on a schedule
that will be driven by the availability of funding
beyond that committed by the TABASGO Foundation.
Phase C is essentially our vision of the ``final''
Network configuration.
Since we do not plan to close down any facilities in the process of Network
development, Phase C facilities include all of Phases A and B, and Phase B
includes all of Phase A.
 
If the funding sought for Phase C should fail to materialize,
the damage to the Network's scientific capabilities would be mostly
a quantitative degradation in number and quality of observations, 
rather than a qualitative change in the kinds of science that could be done. 
Fewer telescopes would mean fewer observations could be carried out per year,
and would also limit scheduling flexibility.
Probably the largest effects would be noticed in the northern hemisphere,
where Phase A/B plans call for few sites and telescopes. Hence
continuous observations would not be possible north of roughly
30$^\circ$ declination, and a few programs that require concurrent
observations could easily saturate the small number of available 1m telescopes.

\subsection{Number and Geographical Distribution of Nodes}

A minimum requirement for a network to be fully capable of time-domain operation
is that it should allow continuous time coverage of objects located anywhere
in the sky.
To achieve this requires
at least three well-separated nodes in each of the northern and 
southern hemispheres.
Thus, from three ideally spaced mid-latitude sites, each separated from the 
others by eight hours of time, night-times overlap by 
about four hours in the hemisphere's winter, and hardly at all in the summer.
Also, northern-hemisphere sites cannot see the southern circumpolar stars,
and conversely.
Therefore, six sites is the minimum feasible for true global coverage.

When choosing potential node sites, we first aimed to keep cost manageable by
considering only sites that are already developed and are occupied by
functioning astronomical facilities.
This not only avoids significant infrastructure development costs, 
but also facilitates
access to trained staff when needed.
Beyond this requirement, the science we wish to do defines our desiderata. 
In order of decreasing importance, these are:  
many good observing hours per year, 
good seeing, dark sky, good prospective scientific collaborations, 
and good internet access.

In practice, this simple and deterministic picture is complicated by geography
(e.g., the northern hemisphere has a wider choice of acceptable astronomical
sites than the southern), by the physical and bureaucratic accessibility
of the sites involved, and by the nature of existing commitments, 
collaborations, and facilities (e.g., LCOGT had operating telescopes in
Australia and Hawaii before we began to plan the 1m network).
With these considerations in mind, we settled on a Phase C configuration of
seven main nodes and one secondary node, 
three in the southern hemisphere and five in the northern
\footnote{http://lcogt.net/network}.
In the southern hemisphere, these are 
Cerro Tololo Inter-American Observatory
in Chile (CTIO), Siding Spring Observatory in Australia (SSO), 
and the South African
Astronomical Observatory at Sutherland, South Africa (SAAO).
The northern (Phase C) main nodes are McDonald Observatory in Texas
(MDO), the Haleakala Observatory on the island of Maui (HO),
the Xinjiang Astronomical Observatory near Urumqi, China (XAO),
and the Teide Observatory on the island of Tenerife, in the Canary Islands (TO).
There are four sites in the north both because of the longitudinal distribution
of suitable sites, and because, though we have a 2m telescope in Hawaii, 
we have not yet succeeded in obtaining permits to place 1m telescopes there.
The secondary node is the Byrne Observatory at Sedgwick (BOS)
in California, near our headquarters;
although we do not expect to make this node a full part of the Network,
it is useful for instrument testing, software development, and 
{\it ad hoc} longitude-specific backup of Network observations.

Table 1 lists the characteristics of these nodes, including the completion status
(Phases denoted PA, PB, or PC) of the telescopes intended for each.
To summarize the table, we have six Phase-A nodes (CTIO, SSO, SAAO, MDO, HO, BOS).
These support 10 Phase-A telescopes:  2 $\times$ 2m telescopes at SSO and HO, 
7 $\times$ 1m telescopes at CTIO, SAAO, and MDO, and the 83cm telescope at BOS.
Phase B will add one additional 1m telescope to MDO, and
2 $\times$ 1m telescopes to SSO.
We anticipate that a total of 11 $\times$ 40cm telescopes will be added at these
sites through 2014 as part of Phase B (see Table 1).
If completely funded, Phase C will add 2 $\times$ 1m telescopes at HO,
and two additional nodes (TO and XAO) including
3 $\times$ 1m telescopes at TO, 2 $\times$ 1m
telescopes at XAO, and about 12 additional 40cm telescopes distributed among 
various nodes.



\begin{deluxetable}{lrrcccllll}
\tabletypesize{\scriptsize}
\tablecaption{LCOGT Network Node Characteristics. \label{tbl-1}}
\tablehead{
\colhead{Node} &    \colhead{\parbox{0.4in}{Latitude\\ (degree)}}    &    \colhead{\parbox{0.4in}{E. Long\\(degree)}}   
& \colhead{\parbox{.3in}{Elev\\ (m)}} & \colhead{\parbox{.4in}{Seeing\\ (arcsec)}} & \colhead{\parbox{0.6in}{Sky Bright\\ (mag $\arcsec^{-2}$)}}  &   \colhead{2m~~~~~~~}   
&   \colhead{1m~~~~~~~~~}   &   \colhead{83cm~~~~}   &   \colhead{40cm~~~~~} }
\startdata
CTIO &-30.1673&-70.8046&2180&0.9&22.0& -- &(PA:3) & -- & (PB:3) \\
MDO & +30.6800 &-104.0151 & 2070 & 1.3 & 22.1  & -- & (PA:1,PB:2) & -- & (PC:3)  \\
BOS &+34.6915&-120.0422 & 360  & 2.5  & 20.7 & -- & -- & (PA:1) & --  \\
HO  &+20.7070 &-156.2575 & 3040 & 1.2  & 22.0  & (PA:1) & (PC:2)  & -- & (PB:2) \\
SSO &-31.2729 & 149.0708 & 1160  & 1.4  & 21.5  & (PA:1)  & (PB:2)  & --  & (PB:2)  \\
XAO &+43.4723 & 87.1760 & 2080 & -- & 21.7 & -- & (PC:2) & --  & (PB:2) \\
SAAO&-32.3806 & 20.8101 &1780 & 1.3 & 22.1 & -- & (PA:3) & -- & (PB:2)  \\
TO  &+28.3004 &-16.5115 & 2390 & --  & --  & -- & (PC:3)  & -- & (PC:3)  \\
\enddata
\tablecomments{
Column descriptions:  (1) node name, (2) north latitude (degrees),
(3) E. Longitude (degrees), (4) elevation (m), (5) median seeing full width
at half maximum ($\arcsec$), (6) typical sky brightness (equivalent V magnitude
per square $\arcsec$), (7) number of 2m telescopes, with development phase
(PA = Phase A, etc.) (8) as column (7), but number of 1m telescopes
(9) as column (7), but number of 83cm telescopes, (10) as column (7), but 
number of
40 cm telescopes.
An entry of ``--'' means ``none'', or ``inadequate information''.
}

\end{deluxetable}


\subsection{Number and Size of Telescopes}

Choosing the size of telescope for the network, and the number of telescopes
per node, involved tradeoffs among the per-telescope cost, the per-telescope
performance, and network system considerations.
A key point proved to be that network scheduling flexibility is greatly
improved by having at least two telescopes at each node.
Moreover, since we expect spectroscopy to be an important part of our
observation mix, and since spectroscopic observations tend to involve long
integration times (and hence are awkward to interrupt), we planned to put
3 telescopes at as many sites as we could afford.
Last, holding collecting area equal, it is less costly to build many
small telescopes than one large one.
One of our earliest activities was thus to design and build a number of
40-cm telescopes.
This was both to validate design concepts to be applied to larger
instruments, and eventually to offload some educational and other
bright-object observations from
our 2m telescopes to smaller, more cost-effective ones.
On the other hand, more telescopes means more components to fail, and
the relative cost of instruments grows as telescope apertures shrink.
And, of course, a certain minimum aperture is required to do the kinds of
science for which the network is intended.

After considering all this (and more), we concluded that the best compromise
was to build 1m telescopes, with the intention of placing three telescopes at
each node.
Recent economic events have made this goal temporarily unreachable,
but as a long-term goal we continue to aim for a Phase C
network having 2 or 3 $\times$ 1m telescopes per node
at each of seven nodes, for a total of 17 $\times$ 1m telescopes.

\section{TELESCOPES}

Our goal is to provide a homogeneous network of robotic telescopes, 
with identical instrumentation, which will allow continuous monitoring of 
astronomical targets from both hemispheres.

We currently operate (Phase A) or plan to operate (Phase B, C) the following facilities: 
\begin{itemize}
\item{The 2m Faulkes Telescopes (FTN at Haleakala, Hawaii, and FTS
at Siding Spring, Australia). (Phase A)}
\item{The 1m Telescope Network, with nodes at sites worldwide (Phase A, B, C).}
\item{The 83cm telescope, at the Byrne Observatory at Sedgwick 
in California. (Phase A)}
\item{The 40cm Telescope Network, also at sites worldwide (Phase B, C).}
\end{itemize}

Optical and mechanical characteristics of each of these telescope classes 
are given in
Table 2; we describe each telescope class in more detail in the 
remainder of this section.



\begin{deluxetable}{lllcccccc}
\tabletypesize{\scriptsize}
\tablecaption{LCOGT Telescope Characteristics.}
\tablehead{
\colhead{Aperture} & \colhead{f/\#} & \colhead{Mount}  &  \colhead{\parbox{0.6in}{Corr. FOV\\ (arcmin)}}   
& \colhead{\parbox{0.6in}{Img. Scale\\ $\arcsec$ pix$^{-1}$}} & \colhead{\parbox{0.65in}{Slew Rate\\ (degree s$^{-1}$)}} & \colhead{Instruments} & \colhead{AGs}
& \colhead{\# Avail. (PC)} }
\startdata
2m & f/10 & Alt/Az & 15 & 0.155 & 2 & 5 & 4 & 2 \\
1m & f/8 & Equatorial  & 46  & 0.386 & 6 & 5 & 1 & 17  \\
83cm & f/8 & Equatorial & 15 & 0.483 & 7  & 3 & 2 & 1  \\
40cm  & f/8 & Equatorial  & 36 & 0.965 & 10 & 2 & 1 & 23 \\
\enddata
\tablecomments{
Column descriptions:  (1) Telescope aperture, (2) Cassegrain focus f/ratio,
(3) Type of telescope mounting, (4) Corrected field of view diameter (arcmin), 
(5) Cassegrain image scale (arcsec per 15 $\mu$m pixel),
(6) Slewing speed (degrees per second),
(7) Number of instrument ports (excluding dedicated autoguiders),
(8) Number of available system autoguiders,
(9) Number of telescopes of this type available for deployment (Phase C).
}

\end{deluxetable}


\subsection{Faulkes 2m Telescopes}
The two Faulkes Telescopes were designed and constructed by 
Telescope Technologies Limited (TTL).
The telescopes feature Ritchey-Chr\'etien Cassegrain f/10 optics,
with solid primary mirrors of Astro-Sital, 20 cm thick,
and secondaries of the same material, on Alt-Az mounts.
We correct
minor astigmatism in the current secondary mirrors using
flexure assemblies driven by compressed-air pistons.
These astigmatic secondaries will soon be replaced with 
lightweighted Astro-Sital mirrors.
Primary mirror supports consist of 36 air-pressure driven support pads,
which are
automatically adjusted to compensate for elevation-dependent
flexure.
The secondary mirrors are adjustable in focus.
Secondary mirror positions are adjusted dynamically (but only when the
instrument shutters are closed) to compensate for changes in telescope
elevation angle and in ambient temperature.

\begin{figure}[H]
   \epsscale{0.75}
    \plotone{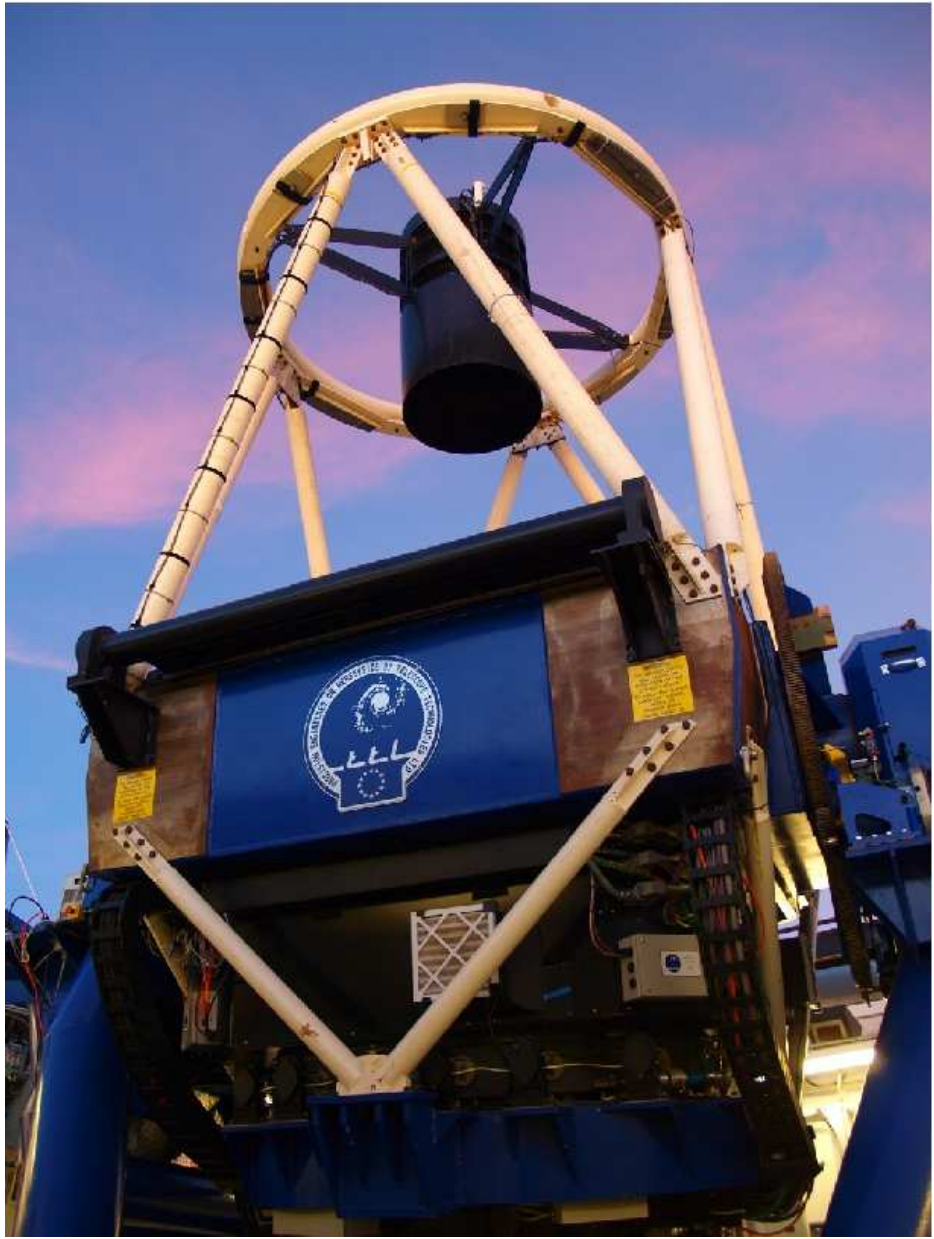}
    \caption{LCOGT's Faulkes North (FTN) 2m telescope at dusk, with the
clamshell enclosure open.  Faulkes South (FTS) is a twin of FTN, but located
at Siding Spring Observatory in Australia.}
    \protect\label{FTN}
\end{figure}

The 2m mountings are Alt-Az designs, with natural frequencies 
in excess of 10 Hz. 
Oil-pad bearings support
both altitude and azimuth motions;
A large ball-bearing-supported Cassegrain
rotator compensates image rotation.
We drive all axes with opposed servo motors
and gear boxes, with feedback provided
by motor encoders and by optical tape encoders attached to the driven parts.
The maximum slewing speed in all axes is 2 degrees per s, and the servo
settling time is less than 5 s.
Thus, pointing to new objects seldom takes longer than 45 s.
Blind pointing is accurate from 3 to 10 arcsec depending on the Cassegrain
axis position.
Periodic tracking errors due to the influence of gear teeth are minimized using correction look-up tables.

Up to five different instruments 
can be mounted at the 
Cassegrain focus, one in the ‘straight through’ position and 
four smaller ones on side ports accessible by
rotating the ‘science fold’ 
tertiary mirror. 
Imaging instruments available on the telescopes are the ``Spectral'' 
(Fairchild Imaging CCD486) and ``Merope'' 
(E2V-4240) optical cameras,
and a Lucky Imaging and High-Speed Imaging (LIHSP) camera. 
We recently deployed on each of the Faulkes telescopes 
a low-resolution cross-dispersed spectrograph (known as FLOYDS) 
with 350 nm to 1100 nm coverage at better than 1.2 nm spectral resolution.
All of these instruments are described in more detail in Section 5.

The 2m telescopes are housed in large (10 m. square) clamshell enclosures,
with shutters operated by hydraulic pistons.
When open, the clamshells allow unvignetted views of the sky for all
elevations greater than 20 degrees above the horizon.
In use, these enclosures have proved to be reliable and leak-free.
They also provide space to deploy additional 40cm telescopes 
on elevated platforms within
the 2m enclosure.

\subsection{The 2m Robotic Control System}

The two Faulkes Telescopes are controlled by a Robotic Control System (RCS) 
\citep{2004SPIE.5493..331F}. 
The RCS runs on a local computer at the site in parallel to the 
Telescope Control System (TCS), which controls the telescope functions such as 
slewing, tracking, autoguiding etc, and the Instrument Control System (ICS), 
which controls the functions of the instruments. 
While the telescope is operating, it is the RCS that issues instructions 
to the TCS and ICS. 
The system features built-in recovery functions to address problems 
automatically, for example an instrument failing to initialize properly 
\citep{2006AN....327..806M}. 
Additionally, the telescopes have local weather stations which 
serve information on humidity levels, cloud cover, precipitation, 
wind speed and temperature to the RCS.
If any of these parameters should exceed their allowed ranges, 
the enclosure is automatically closed and the system goes into stand-by mode.

When atmospheric conditions (seeing, extinction) are too poor 
to allow normal science observations, the RCS switches to observing 
background standard stars. 
These are observed fairly frequently in order to assess changes in 
atmospheric conditions so that normal operation can resume when the 
atmospheric parameters return within their acceptable limits.

For science observations, the RCS receives instructions from 
a database of observations, known as the Phase-II DB. 
This is stored locally at each telescope site and contains all the 
observing programs with their specifications. 
The RCS uses this information to determine the set of instructions to issue 
to the telescope and instrument systems. 
The actual scheduling and submission of the observing requests to 
the telescopes is handled by a separate ``dispatch'' scheduler. 
This runs in real-time; whenever an observation is completed, it lists the
observations in the Phase II DB that are then able to run,
ranks these in order of priority (which is a complex function of many variables,
some of which depend on current observing circumstances),
and dispatches the highest-ranked feasible observation to be executed by the
telescope.

The RCS accepts input from two sources. 
The first one is the Observer Support System (OSS). 
This controls access to the database of observations uploaded by 
the users to the telescope or submitted automatically via external agents. 
The second source of input is the Target of Opportunity Control System (TOCS). 
This is an override program whose execution interrupts the observing schedule 
and initiates immediate observations of a high priority target. 
The TOCS can be started automatically via external triggers 
or may be invoked manually. 
Typical examples of frequent users of this system are responses to alerts 
of Gamma-Ray bursts and anomalous microlensing events.
The OSS can respond to astronomical alerts within 
minutes \citep{2008AN....329..321S},
whereas requests through the TOCS can put a telescope on a new target and
begin observing within tens of seconds.

We are in the process of replacing the software just described with a structure
that is compatable with that for the 1m Network (described below).
This replacement will occur in phases, with the aim of completing the
transition by the end of 2014.
This replacement will maintain almost all of the capabilities of
the old system, while adding valuable new features, notably the ability
to schedule observations that involve coordination between the 2m telescopes
and the 1m Network.

\subsection{Archiving}

The image archiving system for the 2m telescopes is separate from and
predates the 1m system, which is described in section 6.6.
In the 2m system, incoming images are transferred from each telescope
to the central archives 
every ten minutes. 
They are then available on a quick-look page so that, 
if required, an initial assessment of the data quality can be made. 
We calibrate the images hosted on the quick-look pages using 
the most recent flat-fields and biases, usually from the previous night. 
However, since new twilight flats are obtained automatically 
at the end of each night, we recalibrate the images using 
the latest flat-field frames before storing them in the permanent archive.
Data are filed under the proposal name that generated them,
with password-protected access for a proprietary period (normally one year,
for scientific data).
Data taken for educational purposes are made available to the public
immediately.

\subsection{1m Telescopes}

The 1m telescopes in the global Network are designed to fulfill LCOGT's
objective to provide homogeneous, maximally available optical monitoring
of time-variable sources. Each 1m telescope must provide
reliable robotic operation for long periods of time, with minimal
hands-on maintenance, deliver good pointing, tracking and guiding, and
provide uniform image quality. 
We manufactured optical and mechanical parts to tight
tolerances to achieve uniform image scale and
consistent performance among telescopes and instruments.

\begin{figure}[H]
    \epsscale{0.80}
    \plotone{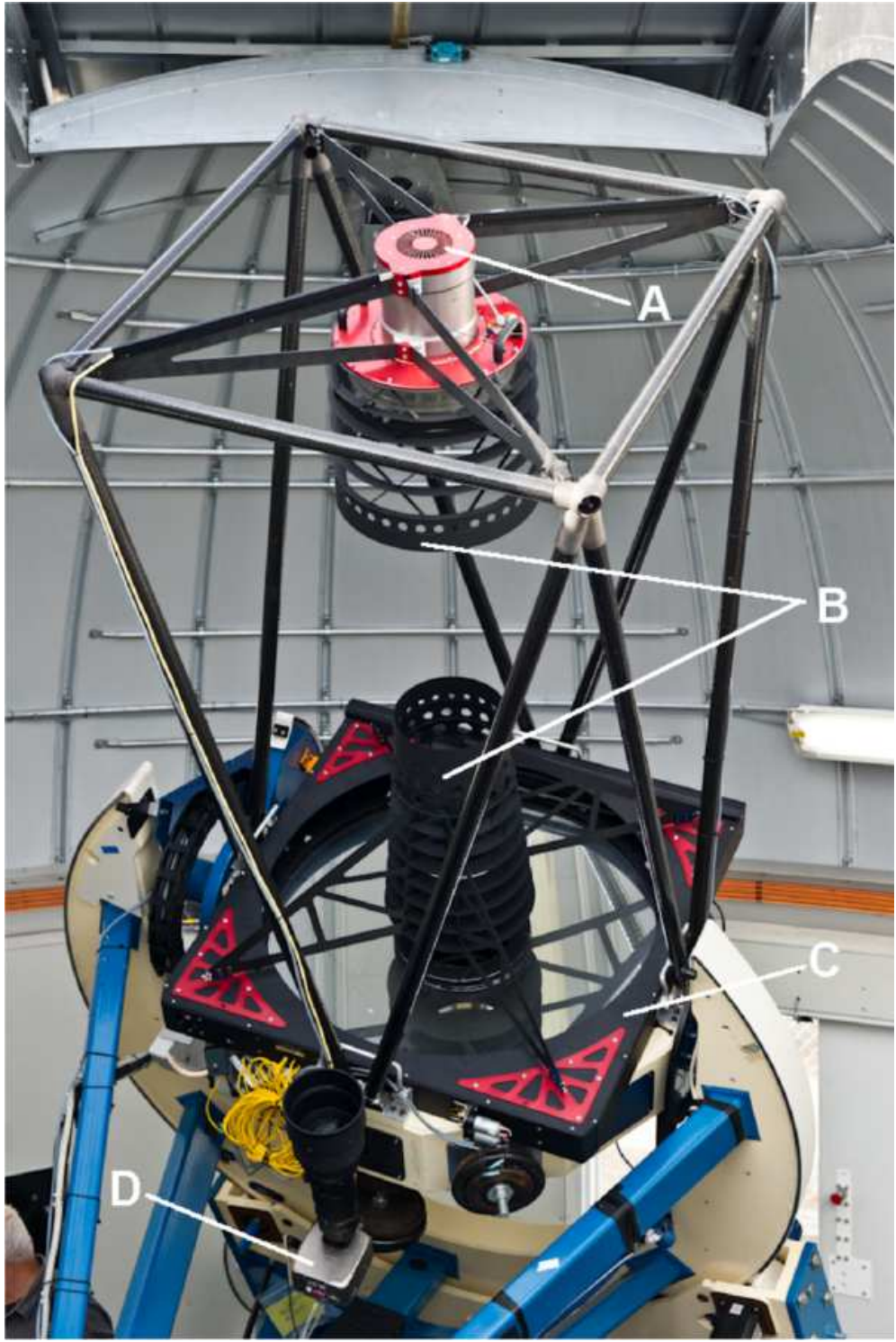}
    \caption{1m telescope, assembled in dome.  
(A) Secondary mirror tip-tilt-focus mechanism.
(B) Lightweight, low-wind-resistance light baffles.
(C) Roller-shade primary mirror cover, with integral Hartmann mask.
(D) Wide-field Extinction Camera.
Photo: Matt Miller, HazardousTaste.com}
    \protect\label{1mtel}
\end{figure}

The reasons for choosing a C-ring equatorial mount are detailed in
\citet{2010SPIE.7739E..55D} but summarized here for completeness. Alt-Az
mounts offer obvious mechanical advantages for larger telescopes, but
dramatically increase the dynamic range requirements for the drive
servo.
As well, an Alt-Az configuration would require the addition of a 
Cassegrain rotator axis under servo
control, resulting in undesirable complexity and weight for the 1m
telescope class. The C-ring concept has long been known to provide
accurate pointing and tracking performance (cf. Palomar 5m, NOAO and AAT 4m),
and we were also influenced by the known long-term robotic reliability of the
2MASS telescopes. C-rings are not popular for visual astronomy because of the
difficulty of human eye access to their focus, or for telescopes
requiring frequent facility instrument changes, but they provide a
stable and reliable mount, with sufficient instrument
clearance for our purposes.
Total instrument depth below the bolt-circle on the back of the primary
mirror cell is 650 mm,  with an overall Optical Tube Assembly (OTA)
length (including instrument) of 2.5 m.

The LCOGT 1m telescopes are modular, each comprising 15 basic structural,
optical and control components, which we ship to site in a few
pre-aligned assemblies. Completed mounts weighing two tons (base,
2 m-diameter C-ring and mirror-cell) are aligned in our shop, shipped to
site, and craned through the 2.1m open shutter aperture of our domes
onto their metal pedestals and concrete piers, with the primary mirror
about 2m above ground level. Each telescope is compact, fitting inside a
6m diameter modified Ash-dome. 
Minimum clearance within the dome
is only about 100 mm in a few locations.
The telescopes can point and track under the pole,
and within 2$^{\circ}$ of the pole,
but the C-ring limits E-W tracking to just over 5 hours from the
meridian, conforming to our specified horizon limit of 15 degrees or 3.7
airmasses.

The OTA design and finite element analysis
details are given in \citet{2010SPIE.7739E..56H} and again summarized
here. We considered a traditional Serrurier truss design, but a central
box section would have required a larger C-ring and heavier OTA and
mount. Instead, we adopted a ``Gemini" like concept in which 
the Mirror cell forms
both the declination axis and the main load bearing structure for the
OTA. The mirror cells are weldments that were each post-machined in one pass
for precise alignment of the instrument bolt circle, and of the optical support
points. The C-ring opening is pinched slightly during assembly and held
together by the mirror cell; these contrasting forces preload the mount,
reducing movement between parts as loads change during
telescope motion.

We designed the optical system, described in \citet{2010SPIE.7739E..56H},
for good image quality, usually limited by
site seeing, and for photometry over a large enough field to provide many
stars for astrometric and relative flux measurements. Our final 1m
optical design comprises an f/2.5 Hextek lightweight primary and 330mm
diameter Hextek secondary, both optically finished by LZOS in Russia,
providing an f/8 modified Ritchey-Chr\'etien
system with the addition of a doublet corrector in front of the
instrument package.
The system is designed for 80\% enclosed energy within a circle
of diameter 0.6 arcsec.  
We paid considerable
attention to optimizing the baffling against stray light from
astronomical sources, such as the moon, to allow for good air flow,
and to minimize ghosting in
the Cassegrain field.

Our lightweight Hextek borosilicate mirrors have an expansion
coefficient of 3.3 ppm C$^{-1}$ at typical operating temperatures.
We employ a
traditional 18-point whiffle tree design for the primary mirror axial support,
and constrain the mirror in the transverse directions using
 a stainless steel hub attached to the upper surface of the
instrument bolt circle, with two slightly springy plastic rings to
press against the inner diameters of the top and bottom
primary glass plates. This results in measured primary movements of
order 70~$\mu$m radially and axially during telescope motion from the
zenith to horizon. But these motions are repeatable, and hence can
be compensated by appropriate terms in the telescope mount model.
We find that hysteresis in these motions is 10~$\mu$m or less (about 4
$\arcsec$ projected on the sky).
To hold the  secondaries without inducing thermal stresses, 
we fabricated powder cast and post-machined central hubs of 
invar tuned to the same CTE as the glass, 
and epoxied into the secondary mirror central holes
 prior to polishing.
We then support the secondary mirror hubs by a
compact 3-axis system 
that provides focus and remote tilt collimation to
sub-micron repeatability. 
We set mirror alignment transverse to the optical axis
during telescope assembly.

Both primary and secondary mirrors move slightly and change their radii
as a result of changing ambient temperature and telescope zenith angle,
which in turn cause variations in telescope focus, collimation, and pointing.
We largely compensate collimation errors by tuning the support system
so that there is substantial cancellation of the primary and secondary 
mirror tilts with zenith angle.
We compensate the remaining collimation errors, as well as errors in
focus, by adjusting the secondary mirror in tip, tilt, and position
along the optical axis.
Focus adjustments take place when they exceed half of the optically 
measurable limit,
and automatically occur between science exposures, when the main instrument shutter
is closed.
We absorb residual errors in pointing into the telescope mount model.

1m telescope pointing is currently about 6 arcsec RMS over the
sky. Good polar alignment is critical to avoid differential image motion
between the science imager and off-axis guiders. This is being achieved
to a few arcsec by a combination of analysis with 
Tpoint software, and flash World Coordinate System (WCS)
fitting via astrometry.net. The 1m telescopes currently slew at 6 
deg s$^{-1}$
(slower when the safety system detects the presence of people in
the dome), and the dome azimuth drives have been upgraded so that the
telescope can move to tracking on any new source in 30 sec or less. Open
loop (unguided) tracking error is about two arcsec per hour. Guided tracking
is accurate to better than 0.5 arcsec or about one pixel.

Our Hextek primaries contain 108 air ``pockets", each about 1-liter in
volume. These are open at the bottom, and tend to trap warm air as the
night cools. This leads to slight ``dimpling" of the surface, visible in
pupil images, and also to significant focus and spherical aberration
concerns due to non-thermalization of the primary. The secondary is
largely immune to this effect as its holes face up and warm air escapes, but we
do provide a fan system drawing air across M2. We were able to overcome
these disturbing effects in M1 by routing a ``sucker" system with a
plastic tube into each of the 108 pockets, continuously sucking the air
out to be replaced by air at ambient temperature.

We cool cryogenic detectors with closed-cycle refrigerators (Cryotiger),
and warmer instruments and electronics with a circulating refrigerated
propylene glycol mixture.
We vent all removed heat directly to the outside of the telescope enclosure.
We maintain the air surrounding the
primary and OTA at ambient temperature by a combination of dome fans
drawing outside air in through the open slit, and mirror cell fans
drawing air down over the primary and out the back of the cell. The
combination of fans and sucker system maintains excellent thermalization
of the primary, removes dimpling, and minimizes residual spherical
aberration. 

\subsection{Byrne Observatory at Sedgwick Telescope}
\label{sec:bos}

The Byrne Observatory at Sedgwick Reserve (BOS) hosts a 0.83m 
telescope built around an
RC Optical Systems\footnote{http://www.rcopticalsystems.com/} 
optical tube assembly. 
The observatory was developed in agreement with UC Santa Barbara 
in the hills near Santa Ynez, CA. 
The land is part of the University of California natural reserve system. 
BOS is located 60 km from LCOGT headquarters, in much darker skies.
It is nevertheless only about 25 km from the Pacific ocean, 
so it frequently suffers from 
high humidity and fog, especially during the summer months.  
The facility is used nightly by LCOGT staff and UCSB students for robotic
observations, and also 
hosts about a dozen star parties per year.

The classical Ritchey-Chr\'{e}tien optical system has a primary mirror
diameter of 83 cm, and a secondary of 29.8 cm.
Working at f/7.97, the focal length is 6.615 m, 
giving a plate scale of 31.5 $\mu$m arcsec$^{-1}$ at the focal plane.  
The equatorial fork style mount was designed and built by LCOGT
in 2007-2008.
Along with its 6 m modified Ash-dome, it was intended as a 
full scale prototype to prove the friction drive, 
motors, and enclosure now used on the 1m telescopes.
We chose a fork mount over a C-ring for the BOS site to allow
eyepiece access for public outreach.
The mount can reach (HA$\pm$5.3 hours, elevation 
\textgreater 20$^{\circ}$), and slews
at a maximum rate of 7.5 degree s$^{-1}$.  
The telescope, instruments, and dome are routinely run under
remote or robotic control.

For two years BOS ran independently of the nascent 1m Network, 
and proved effective as a tool for science, for public outreach,
and for astronomy course support at UCSB.
It has now been upgraded to use
the same mechanisms and software systems as the broader Network.
In this form it will add a new role as testbed for the NRES
spectrograph prototype (see section 5.6.2).

\subsection{40cm Telescopes}

LCOGT's 40cm class of telescopes is 
based on a Meade 16-inch telescope, but with extensive modifications to all
moving parts. 
We have assembled 23 of these telescopes, though in Phase A,
none of these have been permanently deployed to remote sites.
Phase B calls for deploying 11 of them to five nodes, as shown in Table 1.

\begin{figure}[H]
    \epsscale{0.85}
    \plotone{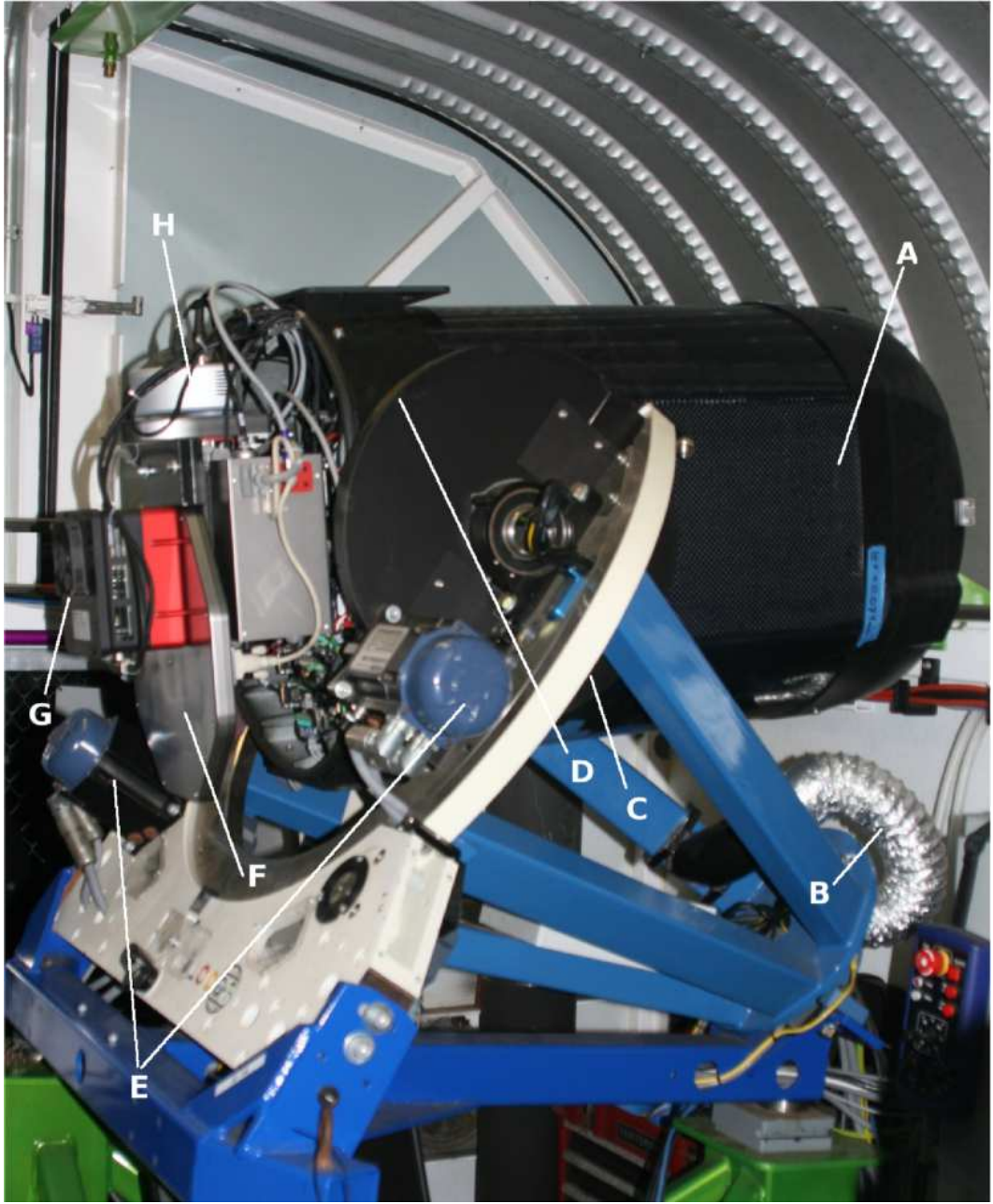}
    \caption{40cm telescope in its Aqawan enclosure.  (A) Carbon-epoxy 
telescope tube.  (B) Duct for optics tube ventilation. (C) Polar axis bearing,
with white fabric dirt guard. (D) Declination axis drive ring.  (E) Direct-drive
servo motors for RA (left) and Declination (right) axes. (F) 8-position
filter wheel for main science camera.  (G) SBIG STX-6303 main science CCD 
camera.  (H) Andor Luca R LIHSP EMCCD camera.}
    \protect\label{40cmtel}
\end{figure}

Our modifications of the telescopes consist of rework on the
OTA to ensure 
stable focus and collimation and to provide thermal
control, and also construction of a custom equatorial mount to carry the OTA.
This mount is a C-ring design
that shares the hardware drive and software 
control mechanisms of our 1m mounts. 
Indeed, we used the 40cm mount development as an opportunity to
prototype the 1m mount, and it proved very useful in clarifying and
providing solutions for problems inherent in this basic design.

The 40cm optical specifications are given in Table 2.
In sky tests of several of these optical systems from our offices near Santa
Barbara, we have found that the telescopes deliver sub-arcsecond imaging
when they are in thermal equilibrium and accurately collimated.
To keep the telescopes close to the ambient temperature, we use a
ducted fan to circulate filtered air from outside the enclosure through
the inside of the OTA.

We will mount most of the Phase-B 40cm telescopes on steel 
piers roughly 1m high,
within small clamshell enclosures that we have dubbed Aqawans.
(``Aqawan'' is a Chumash word meaning ``to keep dry''.)
Each Aqawan has room for two telescopes, placed so that they can move
independently to any location on the sky, with the enclosure roof open or
closed (or partly open), without the telescopes colliding with the
enclosure structure or with each other.
The mountings slew at up to 10 degree s$^{-1}$, settle and begin tracking
within 5 s, and with software mount modeling, achieve blind pointing
accuracy of about 10 arcsec RMS for targets anywhere above the 
telescopes' 15$^{\circ}$ elevation limit.
The C-ring mount design allows sky access anywhere up to $\pm$4.7 hours
from the meridian.
Pointing and tracking below the celestial pole is allowed.

\begin{figure}[H]
    \epsscale{0.95}
    \plotone{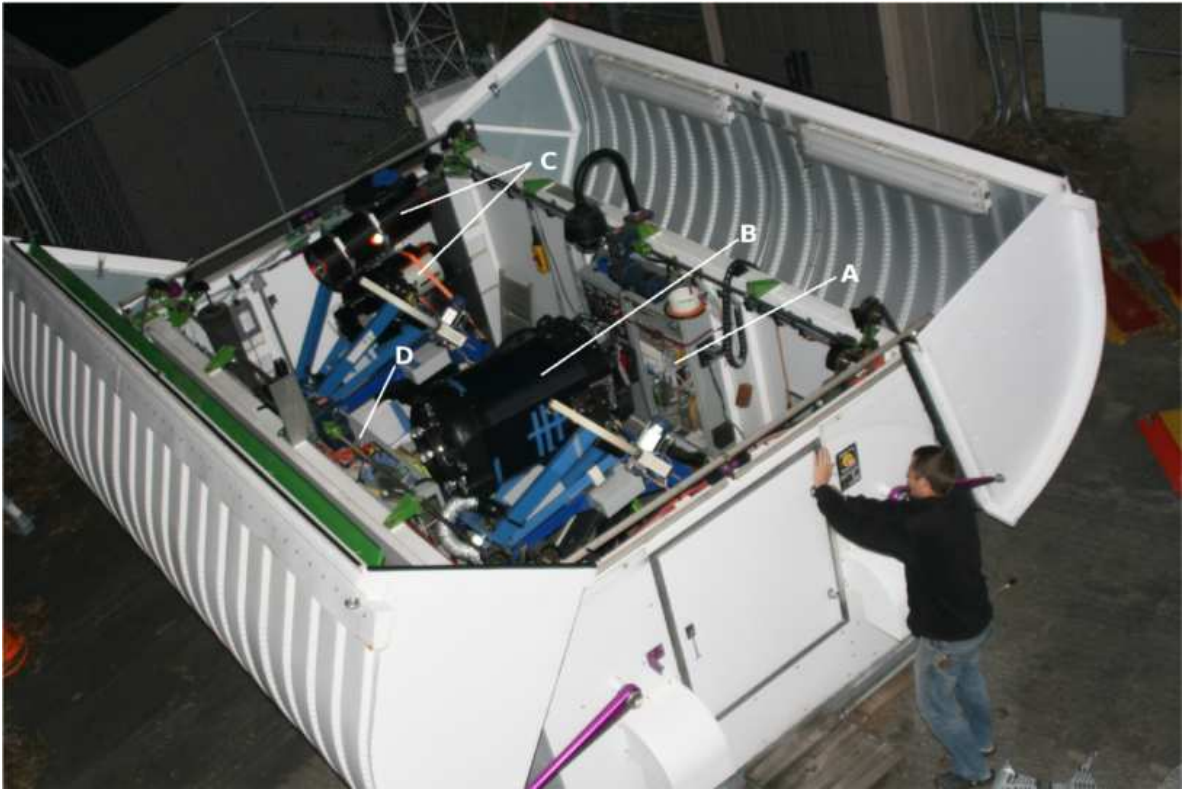}
    \caption{Aqawan enclosure in the fully open position.  (A) Electronics
control system for Aqawan enclosure.  (B) 40cm telescope in C-ring mount.
(C) 20cm astrograph (top) and spectrograph for measuring atmospheric
extinction (bottom) in a single C-ring
mount.  
(D) Electronics control system for two C-ring mounts.}
    \protect\label{aqawan}
\end{figure}

The LCOGT-designed Aqawan enclosures are inspired by but substantially
modified from enclosures built for the MONET telescopes 
\citep{2006SPIE.6270E..55B}.
They are 4.3m long by 2.8m wide, 
and 2.6m tall at the roof
peak.
Their two roof segments have potential to injure people as they open or close,
so we install Aqawans only in fenced and locked areas, and we provide multiple
emergency stop buttons both inside and outside the enclosure.
Control of the enclosures is provided through a custom-designed electrical
panel built around an industrial Programmable Automation Controller.
We provide battery backup for the electrical power needed to open and
(especially) to close the Aqawans, and should this fail, one may open or close
them manually.
We have tested the enclosures through thousands of open-close cycles,
and they display great reliability.
But unfortunately (for this purpose)
our Santa Barbara test facility experiences a limited range of extreme weather.
We therefore anticipate with interest the arrival of the first Aqawan at
a site where snow, ice, and high winds are regular occurrences.

\subsection{Site Environmental Monitoring System (SEMS)}

The core requirements of our environmental system at each site are to
monitor temperature, humidity, dewpoint, wind speed and direction,
wetness, particulates, barometric pressure, and sky brightness 
(solar insolation in
daytime, sky brightness in V mag arcsec$^{-2}$ at night). Temperatures and
humidities are monitored for each site, and at several places for each
telescope, so critical functions like enclosures and mirror covers can
be closed when necessary. 
We extend our monitoring to incorporate data from other 
tenants' site weather instruments,
as well as our own.
In some cases these instruments add information not measured by the SEMS
instrument suite, and even redundant data are useful, since they provide
frequent sanity checks for the SEMS data.
Additionally we provide clean dry air
outlets to our 1m mirrors, that can be activated to prevent dewing.

Our typical limits are that telescope enclosures can be open when the
humidity is less than 90\%, the dewpoint(s) are more than 2 C below
ambient or any mirror temperatures, the windspeed is less than 18 m s$^{-1}$
(65 km h$^{-1}$), particulate count at 1.0 $\mu$m is less than 10$^6$ m$^{-3}$,
and solar zenith distance is greater than 85 $^{\circ}$. The
latter condition allows us to open and thermalize before sunset, stay
open after sunrise, and to capture twilight flat fields, typically
obtained at a point 105$^{\circ}$ away from the Sun, which tends to minimize
the spatial gradients in light from the sky \citep{1996PASP..108..944C}.

We employ a Campbell Scientific CR1000 data logger, with Vaisala
pressure and temperature/humidity sensors, Boltwood cloud sensors,
Unihedron SQM-LE sky brightness monitors, and other sensors. These are
mounted on a 6-m tall tower, typically attached to the side of our
Site Services Building.  At sites such as McDonald Observatory 
that can be affected by
lightning, we also deploy an electric field sensor. The SEMS system is powered
through the site Uninterruptible Power Supply (UPS) to carry it through short power outages. If power remains
off for longer periods, or if the SEMS heartbeat is not detected, our safety system starts a sequenced shutdown
to close all enclosures.

In addition to hard environmental constraints we have adopted softer
constraints such as those based on transparency estimates. We
estimate sky transparency via calibrated Boltwood (sky minus ambient)
temperature sensors and, when we are open, we can also estimate
transparency directly via our science and context cameras, based on
measured magnitudes of known catalog stars.

SEMS data are sampled every 12 s, and persist for seven days. They are then
resampled on a timescale of 120 s, which persists in our database for
2 years. They are then resampled once more on a timescale of 1200 s, which
persists indefinitely. In this way we can match current or historical
data or telescope events to the conditions prevailing at the time.

Additionally we are developing a comprehensive Network Telescope
Operations browser interface, for both internal and external use. This
provides an overview of our Network, summarizes the status of each
site and telescope, and provides graphical monitoring of the
environmental systems described above\footnote{https://telops.lcogt.net}.

\subsection{Safety Systems}

We have implemented a comprehensive set of safety systems into our
global Network. These include interlocks on enclosure doors and gates,
panel doors and ladders (stowed or not) that could interfere with
telescope motion. We have a Fortress Interlocks trap-key system, 
within which keys to
open enclosures or gates must first be removed from a monitored access
panel in the Site Services Building (SSB). This informs the system of the
presence and location of people. Typically the presence of people does
not prevent tracking of affected telescope(s), but does limit
their slew-speed.  If enclosure lights are left on, we automatically 
extinguish them when the
enclosure opens.

In addition to access control,
all telescopes and enclosures have carefully designed systems for the
safety of personnel and equipment. Each enclosure and telescope can be
in one of three states: Automatic (for full robotic control), Manual
for local control only, and Disabled for maintenance. Enclosures cannot
open and telescopes cannot move if deliberately left in manual or
disabled mode.  Our telescopes each have four levels of limit
switches: L1 software limit, L2 electrical sensor position limit (also
used for homing each axis), L3 electrical shutoff, and L4
hardstop. Automatic recovery is possible only from the first two of
these.  Both 2m and 1m telescopes have fail-safe air brake systems; caliper
brakes on both axes are kept off by power and air pressure, and 
come on if either power or pressure fails. The brakes are also turned
on if an error condition occurs.

Our servo systems provide continuous checks for overspeed or
out-of-bounds operation. Our Programmable Automation
Controllers (PACs) provide checks on safe operation for all
sub-systems. Additionally on our 1m Network we have implemented a ``PNOZ''
configurable safety system from Pilz Automation, programmed to detect
and react to a variety of situations. A severe condition may result in
an Emergency Stop (E-stop) being activated, 
which stops and prevents all further motion
on that telescope and enclosure; on-site human interaction is required
to investigate, trouble-shoot and recover from such a situation. If
danger presents, humans can also press red E-stop buttons distributed
around each enclosure and telescope.

Less critical abnormal conditions may result in one or more
components being de-activated, or reset, according to the programmed
reaction. Complex moving interactions such as upper and lower shutters of
our domes are programmed into this safety system: the upper shutter
must open before the lower shutter can start, and the upper shutter
may pause before full closure to first allow the lower shutter to close
completely.

\subsection{Thermal Management}

The Site Services Building contains automatic fans and a large air
conditioning unit to maintain temperature of the building and all
computers near to a 20 C set point. 
The NRES bench spectrograph (see section~\ref{NRES}) 
will be placed within its own
container adjacent to the SSB, and will incorporate even more stable
thermal control than that within the SSB.

Electrical panels within our enclosures are normally ventilated outside the
enclosure with fans. These fans can be automatically turned off or
pulsed in cold, wet conditions, to prevent influx of damp air which
could cause internal dewing. All telescope enclosures are painted white to
reduce heat-load during daytime. Each dome contains a small air
conditioning unit that comes on automatically in hot weather to reduce
internal temperatures and minimize thermalization issues when the
domes open. Each dome has three large wall fans that come on at a
controllable speed when the enclosures open. These draw ambient air
through the open shutter and out of the enclosure, and completely
replace the air in each dome within a few minutes. Thermalization of the
whole telescope environment to ambient temperature typically occurs
within 30 to 60 minutes, depending on temperature differentials
between day and nighttime at the site. As of February 2013 
we have operated in nighttime temperatures in the range from -7 C
(McDonald in the winter) to +30 C (Chile in the summer), but typically
not changing by more than 10 C on any given night at any site. We have
designed for an operating range from -15 C to +40 C.

Each enclosure contains a circulating glycol cooling system,
maintained at a temperature exceeding the local dewpoint, used to
circulate heat away from thermoelectrically cooled cameras, guiders
and electronics crates. The heat from this system is exhausted outside
each enclosure with fans. 1m mirror cells contain eight fans that draw air
over the mirrors and out the back of each cell. The sucker system fan
comes on when each dome opens to thermalize each 1m primary, again
typically within about 30 minutes.

The LCOGT-designed Sinistro camera (covered in detail in the next section)
is cooled to -100 C with recirculating gas from a
Cryotiger unit in a cryo-cabinet inside each dome. The hoses for this
system and the glycol system run through the RA and DEC energy chains
to the primary mirror cell.


\section{INSTRUMENTS}

LCOGT's instrument complement is designed to support a range of science 
goals in time-domain astrophysics. 
Our aim is that every telescope offer a standardized suite of instruments kept
as homogeneous as possible for each class of telescope, 
across all nodes in the network.  
This enables observations to be
carried out promptly by the best available node, or for 
time series observations of a given target to be carried out by a sequence of telescopes around the globe, as night falls at each node. 

LCOGT instrumentation falls into two categories: 
imagers 
and spectrographs.
Our software control system allows any imager to be used 
as an autoguider, including self autoguiding. 
Instruments designed specifically for autoguiding 
(these include facility autoguiders dubbed ``4ag'' and ``2m0'' 
that are mounted on the 2m telescopes, 
and the ``1m0'' autoguider for the 1m telescopes) have independent 
focus control.
For bright-star photometry, this allows in-focus images 
on the autoguider and simultaneous 
defocused images on the science camera. 
By design, the autoguider mounts minimize flexure relative 
to the primary instrument's image plane, 
and have been verified to be stiff against gravity deflection
to about 0.1 resolution element.

Table 3 lists the imaging instruments that are installed on the various
LCOGT telescopes.
Although homogeneity among imagers is highly desirable, this desire is
often outweighed by the need to accommodate differing image scales
among telescopes, differing readout requirements and fields of view
among main science cameras,
high-speed cameras, and autoguiders, and the necessity to deploy telescopes
promptly, even though the imagers ultimately intended for them 
may not yet be ready.
For these reasons, the Network employs a considerable variety of CCD imagers. 
In the remainder of this section, we describe these instruments in more detail.

\begin{deluxetable} {lccccl}
\tabletypesize{\scriptsize}
\tablecaption{LCOGT Network Imager Characteristics. \label{tbl-3}}
\tablehead{
\colhead{\parbox{1.4in}{Instrument name\\ \phm{\quad} Camera type\\ \phm{\quad} Detector type}} & \colhead{\parbox{1in}{\centering Detector format\\ Plate scale}} & \colhead{$\mathrm{QE}_\mathrm{max}$} & \colhead{\parbox{0.4in}{\centering Readout\\ (s)}} & \colhead{\parbox{0.2in}{\centering $m(1e)$\\ ($r^\prime$)}} & \colhead{Filters}\\\vspace{-0.1in}
}
\startdata

\parbox{1.3in}{Merope (2.0)\\ \phm{\quad}Merope\\ \phm{\quad}e2v CCD42-40 DD, BI} &  \parbox{1.5in}{\centering 2048 $\times$ 2048 $\times$ 13.5\\ 4$\arcmin$.74 @ 0$\arcsec$.278 (2x2)} & 90\% & 14 & 24.6 &  \parbox{1.3in}{$u^\prime$ $g^\prime$ $r^\prime$ $i^\prime$ $z_s$ $Y$ $B$ $V$ $R_C$ $I_J$ $\mathrm{H\alpha}$ $\mathrm{H\beta}$ [O III] DDO51 $V$+$R$ ND2 $V_s$} \\
\vspace{0in}\\

\parbox{1.3in}{ Spectral (2.0)\\ \phm{\quad} Spectral 600 \\ \phm{\quad} FI CCD486 BI } &  \parbox{1.5in}{\centering 4096 $\times$ 4097 $\times$ 15.0 \\ 10$\arcmin$.5 @ 0$\arcsec$.309 (2x2) } & 90\% & 11 & 24.6 &  \parbox{1.3in}{ $u^\prime$ $g^\prime$ $r^\prime$ $i^\prime$ $z_s$ $Y$ $UV$ $B$ $V$ $R_C$ $I_C$ $\mathrm{H\alpha}$ $\mathrm{H\beta}$ [O III] DDO51 $V$+$R$ ND2 $V_s$} \\
\vspace{0in}\\

\parbox{1.4in}{ LIHSP (2.0)\\ \phm{\quad} Andor iXon 888 \\ \phm{\quad} e2v CCD201 BI, FT, EM } &  \parbox{1.5in}{\centering 1024 $\times$ 1024 $\times$ 13.0 \\ 2$\arcmin$.29 @ 0$\arcsec$.134 (1x1) } & 90\% & 0.13 & 23.9 est. &  \parbox{1.3in}{ u' g' r' i' $z_s$ Y B V $R_C$ $I_J$ Ha Hb [O III] DDO51 V+R ND2 $V_s$} \\
\vspace{0in}\\

\parbox{1.3in}{ SBIG (1.0) \\ \phm{\quad} SBIG STX-16803 \\ \phm{\quad} Kodak KAF-16803 FI } &  \parbox{1.5in}{\centering 4096 $\times$ 4096 $\times$ 9.0 \\ 15$\arcmin$.8 @ 0$\arcsec$.464 (2x2) } & 50\% & 12 & 23.0 &  \parbox{1.3in}{ $u^\prime$ $g^\prime$ $r^\prime$ $i^\prime$ $z_s$ $Y$ $w$ $UV$ $Bu$ $V$ $R_C$ $I_C$} \\
\vspace{0in}\\

\parbox{1.3in}{ Sinistro (1.0) \\ \phm{\quad} Sinistro (LCOGT) \\ \phm{\quad} FI CCD486 BI } &  \parbox{1.5in}{\centering 4096 $\times$ 4097 $\times$ 15.0 \\ 26$\arcmin$.4 @ 0$\arcsec$.387 (1x1) } & 90\% & 4 & 23.5 est. &  \parbox{1.3in}{ $u^\prime$ $g^\prime$ $r^\prime$ $i^\prime$ $z_s$ $Y$ $w$ $UV$ $Bu$ $V$ $R_C$ $I_C$} \\
\vspace{0in}\\

\parbox{1.3in}{ Autoguider (1.0)  \\ \phm{\quad} FLI ML4720 \\ \phm{\quad} e2v CCD47-20 BI, FT } &  \parbox{1.5in}{\centering 1024 $\times$ 1024 $\times$ 13.0 \\ 5$\arcmin$.72 @ 0$\arcsec$.335 (1x1) } & 90\% & 1 & 23.8 &  \parbox{1.3in}{ $g^\prime$ $r^\prime$ $i^\prime$ $z_s$ $Bu$ $V$ Clear} \\
\vspace{0in}\\

\parbox{1.3in}{ SciCam (0.4) \\ \phm{\quad} SBIG STX-6303 \\ \phm{\quad} Kodak KAF-6303E FI } &  \parbox{1.5in}{\centering 3072 $\times$ 2048 $\times$ 9.0 \\ 29$\arcmin$.7 $\times$ 19$\arcmin$.8 @ 0$\arcsec$.580 (1x1) } & 50\% & 12 & 22.1 &  \parbox{1.3in}{ $u^\prime$ $g^\prime$ $r^\prime$ $i^\prime$ $z_s$ $w$ $Bu$ $V$ } \\
\vspace{0in}\\

\parbox{1.3in}{ SBIG (0.8) \\ \phm{\quad} SBIG STL-6303E \\ \phm{\quad} Kodak KAF-6303E FI } &  \parbox{1.5in}{\centering 3072 $\times$ 2048 $\times$ 9.0 \\ 14$\arcmin$.8 $\times$ 9$\arcmin$.90 @ 0$\arcsec$.580 (2x2) } & 50\% & 20 & 23.2 &  \parbox{1.3in}{B V $g^\prime$ $r^\prime$ $i^\prime$ $z^\prime$ H$\alpha$ Clear } \\
\vspace{0in}\\

\parbox{1.3in}{ Merope (0.8) \\ \phm{\quad} Merope \\ \phm{\quad} e2v CCD42-40 DD, BI } &  \parbox{1.5in}{\centering 2048 $\times$ 2048 $\times$ 13.5 \\ 14$\arcmin$.8 @ 0$\arcsec$.435 (1x1) } & 90\% & 14 & 23.8 est. &  \parbox{1.3in}{ $u^\prime$ $g^\prime$ $r^\prime$ $i^\prime$ $z_s$ $Y$ $B$ $V$ $R_C$ $I_J$ $\mathrm{H\alpha}$ $\mathrm{H\beta}$ [O III] DDO51 $V$+$R$ ND2 $V_s$ } \\
\vspace{0in}\\


\parbox{1.5in}{ LIHSP (0.8) \\ \phm{\quad} Andor Luca R  \\ \phm{\quad} TI TX285SPD-B0 BI, EM } &  \parbox{1.5in}{\centering 1004 $\times$ 1002 $\times$ 8.0 \\ 4$\arcmin$.29 @ 0$\arcsec$.257 (1x1) } & 50\% & 0.13 & 21.7 est. &  \parbox{1.3in}{  $g^\prime$ $r^\prime$ $i^\prime$ $z_s$ $Bu$ $V$ Clear } \\
\vspace{0in}\\

\parbox{1.5in}{ LIHSP (0.4) \\ \phm{\quad} Andor Luca R \\ \phm{\quad} TI TX285SPD-B0 BI, EM } &  \parbox{1.5in}{\centering 1004 $\times$ 1002 $\times$ 8.0 \\ 8$\arcmin$.60 @ 0$\arcsec$.515 (1x1) } & 50\% & 0.13 & 20.1 est. &  \parbox{1.3in}{ $g^\prime$ $r^\prime$ $i^\prime$ $z_s$ $Bu$ $V$ Clear } \\
\vspace{0in}\\

\parbox{1.4in}{ LIHSP (1.0) \\ \phm{\quad} Andor iXon 888 \\ \phm{\quad} e2v CCD201 BI, FT, EM } &  \parbox{1.5in}{\centering 1024 $\times$ 1024 $\times$ 13.0 \\ 5$\arcmin$.72 @ 0$\arcsec$.335 (1x1) } & 90\% & 0.13 & 22.1 est. &  \parbox{1.3in}{ $g^\prime$ $r^\prime$ $i^\prime$ $z_s$ $Bu$ $V$ Clear } \\
\vspace{0in}\\

\parbox{1.4in}{ ExtCam (1.0) \\ \phm{\quad} SBIG STL-6303E \\ \phm{\quad} Kodak KAF-6303E FI } &  \parbox{1.5in}{\centering 3072 $\times$ 2048 $\times$ 9.0 \\ 238$\arcmin$ $\times$ 158$\arcmin$ @ 4$\arcsec$.64 (1x1) } & 50\% & 10 & 17.7 est.  &  \parbox{1.3in}{ B V $r^\prime$ $i^\prime$ $z_s$ } \\
\vspace{0in}\\


\enddata
\tablecomments{
Column descriptions:  (1) Instrument name, (telescope aperture in m),
generic type of dewar/readout electronics, manufacturer's designation
of detector chip.
(2) Detector format shown as (X-dimension) $\times$ (Y-dimension) $\times$ (pixel size in
$\mu$m);  Plate scale shows field of view in arcmin, projected pixel size
in arcsec at the indicated binning (e.g., 2x2).
(3) Maximum detector quantum efficiency (percent).
(4) Full image readout time, at the binning shown in column (2).  
(5) Stellar magnitude in $r^{\prime}$ producing 1 photoelectron per s.
(6) List of filters normally mounted on the imager.
}

\end{deluxetable}

\subsection{Sinistro}
\begin{figure}[H] 
    \epsscale{0.95}
    \plotone{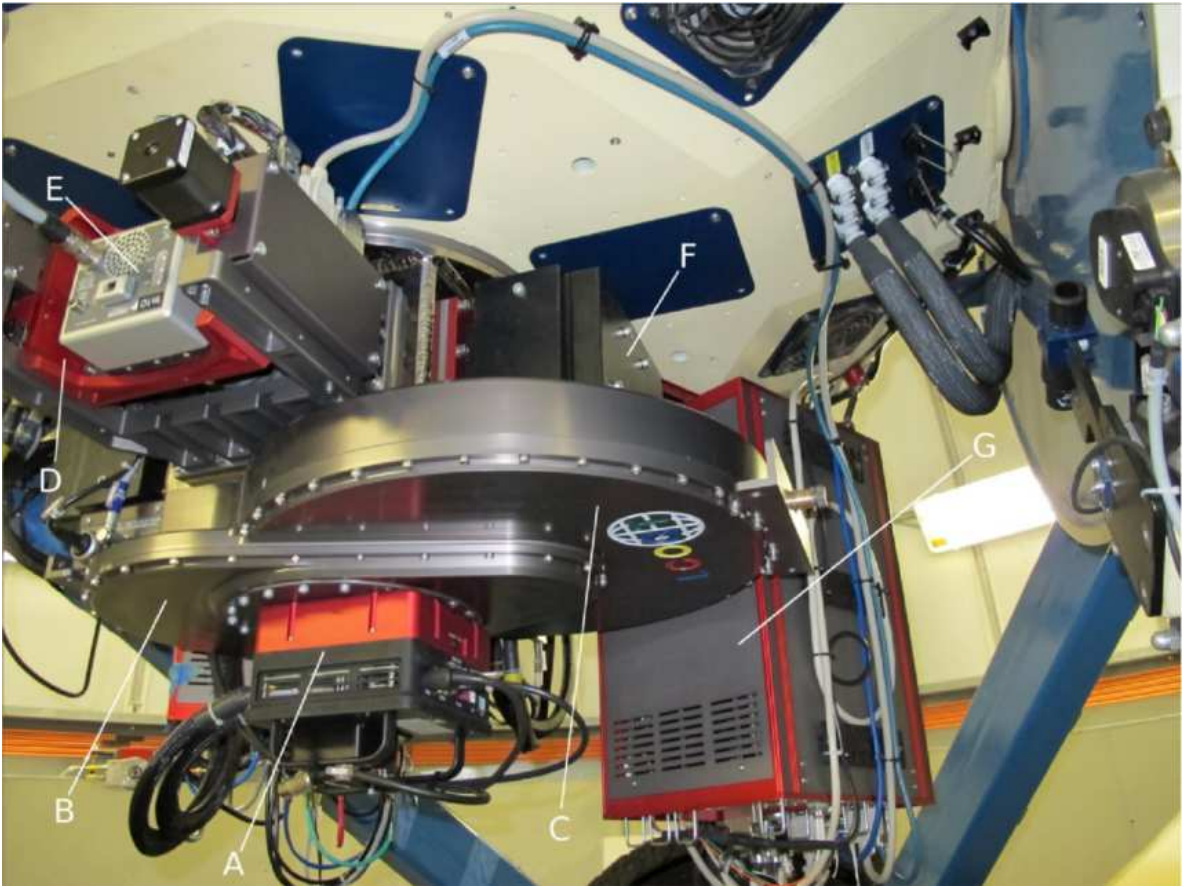}
    \caption{Sinistro instrument package mounted on the tailpiece of a
1m telescope.  
(A) SBIG STX-16803 4K $\times$ 4K CCD camera.  
(B) Rotating-disk shutter assembly.  
(C) 3-layer filter wheel assembly.  
(D) Side port with independent focus stage for LIHSP camera or autoguider.
(E) Andor Luca S LIHSP camera; may also be used as autoguider.
(F) Side port (one of 4) with blank-off panel.
(G) Electronics crate for tailpiece power distribution, signal conditioning,
and related functions.
}
    \protect\label{sinistro}
\end{figure}

\protect\label{sinistro0}
Sinistro is the standard multi-instrument package which will be mounted 
on all 1m telescopes (Fig.~\ref{sinistro}).  
Corrector optics offer a straight-through unobstructed and unvignetted 
116 mm diameter field to the primary science imager. 
As an interim measure for the 1m telescopes we have shipped to date, 
the CCD camera attached to this imager is an 
SBIG STX-16803, a frontside illuminated 4K x 4K device with 9 $\mu$m pixels
giving a field of view 15$\arcmin$.8 arcmin square (see Table 3).
This camera uses a Peltier cooler, and operates at -20 C.

The interim Sinistro CCD camera will soon (Phase B) be replaced with
a Fairchild Imaging CCD 486 BI 4096$\times$4096 device with 
15$\mu$m pixels. 
This is run by an LCOGT designed and built detector controller, 
which achieves a 4\,Mpix s$^{-1}$
readout with a read noise of $\sim$10\,e- pix$^{-1}$.   
The image scale is 25.$\arcsec$8 mm$^{-1}$ or 0.$\arcsec$387 pix$^{-1}$, giving a field of view
26$\arcmin$.6 square.
The instrument is maintained at -100$^{\circ}$C in 
an LCOGT-designed cyrostat, cooled by a Brooks Automation PCC Cryotiger.  
Sinistro's LCOGT-designed filter system comprises three independent 
overlapping wheels, each holding seven square, 75\,mm filters, 3--8\,mm thick.  
This system can change filters within 5\,s. 
and provides 21 usable filter slots.
We list the filters available in LCOGT imagers in Table 4;
in addition to the filters shown there, Sinistro wheels each have a pinhole for optical testing,
and two spare slots. 
The 1m telescope focus automatically adjusts as required by filter changes,
with the focus offsets specified in configuration files, 
but most filters are in fact parfocal.

Sinistro also offers four off-axis ports that, in the absence of seeing,
would produce diffraction-limited images.
The first supports a dedicated, independently focusable autoguider based on 
a high-QE back-illuminated frame-transfer CCD (e2V 4720).
The second houses an independently focusable high speed camera described in
section~\ref{LIimaging}.  
The third is connected to the fiber feed to the NRES spectrograph
described in section~\ref{NRES}, while the fourth port is open for future 
instruments.
These side ports use off-axis regions of the 1m telescope's field,
so that deployable turning mirrors are not required.
Since all instruments view the sky all the time, it is in principle
possible to use multiple instruments simultaneously;
this capability is currently implemented only for guiding during science
exposures.

\subsection{BOS Imaging Instruments}

BOS is outfitted for traditional CCD imaging, as well as high speed
photometry / lucky imaging.
The telescope's primary science camera is a Santa Barbara Instrument Group
(SBIG\footnote{http://www.sbig.com/}) STL-6303E with a 3072x2048 Kodak
KAF-6303E CCD,
TEC+glycol chilled to -20 C,
and mounted to an SBIG filter wheel equipped with filters as listed in Table 4.
We will soon (Phase B) replace this camera with a copy of the Merope
instrument used on the 2m telescopes, with properties listed in Table 3.
The BOS primary science camera is
supported by two unfiltered off-axis autoguiders with
independent focus stages to allow for intentionally defocused observations.  

An Andor Technology\footnote{http://www.andor.com/} Luca R camera is used
for high speed photometry, lucky imaging, and speckle interferometry.
The 1004x1002x8$\mu$m EMCCD can image at 12.4 Hz (full frame).
The focal length is boosted with a 2x Barlow lens for a pixel scale of
$0\farcs{1286}$ pixel$^{-1}$ and a $2\farcm{1}\times2\farcm{1}$ field of view.
This camera can provide diffraction-limited imaging in the optical even
under poor seeing conditions, and high-speed photometry of bright sources.  
The Andor camera can quickly and easily be removed and replaced with
an eyepiece for visual observing during star parties.  
Due to its close proximity to LCOGT's headquarters near Santa Barbara, CA,
BOS is a testbed for prototype instruments and now hosts the prototype
of LCOGT's NRES echelle spectrograph.  


\subsection{Filter Selection}

Our aim is to employ a range of filters spanning the near-UV ($\sim$320nm) 
to near-IR ($\sim$1000nm) which exploit the wavelength-sensitivity of our 
detectors to achieve our science goals.  
We chose broadband filters from two photometric systems: Landolt
(Johnson/Cousins) and SDSS (Sloan primed) owing to their widespread use 
in astronomy.  
After careful consideration of the glass, coatings and exact prescriptions 
used by various manufacturers, we selected Astrodon filters as providing 
the closest match to the original filter response functions of each 
photometric standard.   
On delivery of each filter, we measure the true transmission curve 
and store it in a database.   
To use the strong red sensitivity of our CCDs we will also have 
broadband Pan-STARRS-Z$_{S}$ and -Y filters.
For science requiring high throughput, such as asteroid imaging, 
we will also have the Pan-STARRS-w filter.  
Table 4 lists the names and characteristics of filters that may be found 
on LCOGT instruments.
The LCOGT webpage contains links to the filter transmission curves, CCD quantum
efficiency $vs$ wavelength, and measured mirror reflectance $vs$ wavelength.

\begin{deluxetable}{llcccllll}
\tabletypesize{\scriptsize}
\tablecaption{LCOGT Network Filter Characteristics. \label{tbl-4}}
\tablehead{
\colhead{Type} &  \colhead{Designation} & \colhead{$\lambda_{mid}$} & \colhead{$\delta\lambda$} & \colhead{T$_{\rm max}$} & \colhead{Mfr} & \colhead{Instruments} & \colhead{Notes} 
}
\startdata
\multicolumn{8}{l}{Johnson/Cousins/Landolt/Bessell}\\
\quad $U$ & uv & \phd369.7 & \phn47.2 & 0.975 & Astrodon & Spectral, 1m0 & \\
\quad $B$ & bu & \phd426.1 & \phn88.7 & 0.989 & Astrodon & 1m0, 0m4 & \\
\quad $B$ & bx &  &  &  & Omega & Merope, LIHSP 2m0 & Bessell prescription\\
\quad $B$ & bx & \phd432.6 & \phn98.9 & 0.686 & Cust. Sci. & Spectral & Bessell prescription\\
\quad $V$ & vx & \phd538.7 & \phn96.4 & 0.984 & Astrodon & 1m0, 0m4 & \\
\quad $V$ & vx &  &  &  & Omega & Merope, LIHSP 2m0 & Bessell prescription\\
\quad $V$ & vx & \phd532.5 & \phn83.8 & 0.863 & Cust. Sci. & Spectral & Bessell prescription\\
\quad $R$ & rs & \phd641.3 & 142.5 & 0.994 & Astrodon & 1m0 & \\
\quad $R$ & rc &  &  &  & Omega & Merope, LIHSP 2m0 & Bessell prescription\\
\quad $R$ & rc & \phd628.7 & 122.2 & 0.795 & Cust. Sci. & Spectral & Bessell prescription\\
\quad $I$ & ic & \phd807.1 & 162.3 & 1.000 & Astrodon & 1m0 & \\
\quad $I$ & ic &  &  &  & Omega & Merope, LIHSP 2m0 & Bessell prescription\\
\quad $I$ & ic & \phd877.2 & 304.0 & 0.930 & Cust. Sci. & Spectral & Bessell prescription\\
\quad Clear & clear & \phd700\phd\phn & 700\phd\phn & 0.980 & Astrodon &  & \\
\quad $V$+$R$ & solr-vr & \phd596.6 & 198.8 &  & Asahi & Merope, Spectral, LIHSP 2m0 & \\
\multicolumn{8}{l}{SDSS} \\
\quad $u^\prime$ & up & \phd348.8 & \phn66.5 & 0.80\phd & Asahi & Merope, Spectral, LIHSP 2m0 & \\
\quad $u^\prime$ & up & \phd352.2 & \phn64.4 & 0.99\phd & Astrodon & 1m0, 0m4 & \\
\quad $g^\prime$ & gp & \phd474.0 & 151.0 & 0.93\phd & Asahi & Merope, Spectral, LIHSP 2m0 & \\
\quad $g^\prime$ & gp & \phd475.6 & 147.0 & 0.99\phd & Astrodon & 1m0, 0m4 & Excludes 400nm, 5577 OI \\
\quad $r^\prime$ & rp & \phd622.0 & 134.0 & 0.95\phd & Asahi & Merope, Spectral, LIHSP 2m0 & \\
\quad $r^\prime$ & rp & \phd627.4 & 131.0 & 0.99\phd & Astrodon & 1m0, 0m4 & Excludes 5577 OI \\
\quad $i^\prime$ & ip & \phd765.3 & 138.5 & 0.96\phd & Asahi & Merope, Spectral, LIHSP 2m0 & \\
\quad $i^\prime$ & ip & \phd769.5 & 150.0 & 0.99\phd & Astrodon & 1m0, 0m4 & \\
\quad $z^\prime$ & zp &  &  & 0.998 & Astrodon & Unused, No red cutoff & $>$825.9, no red cut\\
\quad $z_s$ & zs & \phd869.3 & 102.5 & 0.97\phd & Asahi & Merope, Spectral, LIHSP 2m0 & Pan-STARRS cut\\
\quad $z_s$ & zs & \phd873.0 & \phn94.3 & 0.99\phd & Astrodon & 1m0, 0m4 & Pan-STARRS cut\\
\quad $Y$ & yx & 1005.3 & 107.5 & 0.95\phd & Asahi & Merope, Spectral, LIHSP 2m0 & Pan-STARRS\\
\quad $Y$ & yx & 1004.4 & 157.0 & 0.99\phd & Astrodon & 1m0, 0m4 & Pan-STARRS\\
\quad $w$ & wx & \phd623.3 & 443\phd\phn & 0.98\phd & Astrodon & 1m0, 0m4 & Pan-STARRS (g+r+i)\\
\multicolumn{8}{l}{Narrow band}\\
\quad H$\alpha$ & ha & \phd656.0 & \phn\phn9.2 & 0.96\phd & Asahi & Merope, Spectral, LIHSP 2m0 & \\
\quad H$\alpha$ & ha & \phd656.1 & \phn\phn3.0 & 0.95\phd & Astrodon & 1m0 & \\
\quad H$\beta$ & hb & \phd486.1 & \phn\phn5.0 & 0.82\phd & Asahi & Merope, Spectral, LIHSP 2m0 & \\
\quad [O III] & o3 & \phd501.6 & \phn\phn6.0 & 0.85\phd & Asahi & Merope, Spectral, LIHSP 2m0 & \\
\multicolumn{8}{l}{Skymapper}\\
\quad $Vs$ & vs & \phd383.6 & \phn27.6 &  & Asahi & Merope, Spectral, LIHSP 2m0 & Skymaper CaII $V_s$ filter\\
\multicolumn{8}{l}{DDO (David Dunlap Observatory, McClure and van den Bergh)}\\
\quad 51 & ddox-51 & \phd515.1 & \phn15.4 & 0.89\phd & Asahi & Merope, Spectral, LIHSP 2m0 & \\
\enddata
\tablecomments{
Column descriptions:  (1) Filter series and name, (2) LCOGT internal designation, (3) Central wavelength (nm), (4) Spectral bandwidth (nm), (5) Peak transmission,  (6) Manufacturer, (7) Instruments using filter, (8) Comments
}

\end{deluxetable}


\subsection{Lucky Imaging and High-Speed Photometers}

Several of LCOGT's scientific goals require occasional high-resolution
imaging.
Examples include photometry of microlensed stars in very dense starfields in the
galactic bulge, 
searches for background eclipsing binaries in the close neighborhood of
suspected transiting-planet candidates,
and detection and characterization of binary asteroids.
Moreover, occultation studies of solar system objects and eclipses of
compact stars demand high temporal sampling rates.
Unfortunately, present adaptive-optics sytems remain too expensive for us
to deploy to all of our sites.
We have therefore chosen to install Electron Multiplying CCD (EMCCD) detectors
on both of the Faulkes Telescopes
and on four of our 1m telescopes (one at each of four sites).
These enable high speed imaging for 
both high speed photometry and resolution enhancement techniques 
(e.g., ``lucky imaging''). 
We refer to these instruments generically as LIHSP: 
Lucky Imaging and High Speed Photometers. 

High cadence imaging permits photometry of events that require short 
time -- typically subsecond -- monitoring
(e.g., occultations \citep{1979ARA&A..17..445E}). 
Moreover, the ability to collect rapid measurements of astronomical 
targets allows a variety of resolution enhancing techniques: lucky imaging 
\citep{2006A&A...446..739L, 2012PASP..124..861G}, speckle interferometry 
\citep{1983ApOpt..22.4028L,2007A&A...475..771S}, 
and image deconvolution \citep{2011A&A...531A...9H} 
all require short exposure ($>$10~Hz, on bright targets, typically R$<$14 
for the 2m and R$<$12 for our 1m telescopes) to achieve spatial resolution 
beyond the limitations imposed by the atmospheric effects of image smearing. 

All of the techniques mentioned above are implemented at LCOGT, where we
have achieved a spatial
resolution of 0.3 $\arcsec$ at the Faulkes Telescopes. 
Although lucky imaging techniques generally should allow diffraction
limited imaging, the 2m optical systems are not diffraction limited.
Although Barlow lenses decreasing the plate scale may achieve
better than 0.3 $\arcsec$ resolution, we do not expect to be able to
reach the diffraction limit
on the 2m telescopes.
Results may be better for our smaller-aperture systems;
we expect heavy use of the 40cm telescopes to produce high-resolution
images for education and outreach.

For the 2m and 1m telescopes, the high speed 
imaging camera is an Andor iXon 888, using a back illuminated EMCCD, 
while many of the 40~cm telescopes will host an Andor iXon Luca R 
with a front-illuminated EMCCD.

In EMCCDs, a solid state Electron Multiplying (EM) register added to the 
end of the normal serial register amplifies the signal before the
readout amplifier introduces 
noise. 
Thus EMCCDs allow the detection of signal in photon starved conditions. 
A particularly useful application in astronomy is high cadence imaging.  
The cameras installed on our 2m and 1m telescopes allow imaging at a 
cadence as high as 8~Hz on the full field: 2$\arcmin$.3 square at the FTs, 
and 5$\arcmin$.7 square at the 1~m telescopes, with a pixel resolution of 
0$\arcsec$.13 pix$^{-1}$ and 0.$\arcsec$34 pix$^{-1}$ respectively. 
On the 40~cm telescopes the Luca~R allows full field imaging at 
12.4~Hz cadence, with a field of view measuring 4$\arcmin$.3 square, 
and a plate scale of 0.$\arcsec$26 pix$^{-1}$. 
The final image scale and field of view is still to be determined for all
telescopes, based on science needs, and may be modified with the use
of Barlow lenses.
Experience suggests that on the 2m Faulkes Telescopes, a scale of 0$\arcsec$.07
pix$^{-1}$ may be better.
Subframing and binning allow us to achieve  higher cadences: tens, or even 
hundreds of Hz. 
We typically operate the detectors in frame transfer mode,
so the duty cycle for each exposure cycle is 
$>$ 99\% at 8 Hz, and still $>$~50\% at the 
highest cadences available.

\protect\label{LIimaging}

\subsection{FLOYDS}
\protect\label{FLOYDS}

The FLOYDS instruments are a pair of nearly identical, low dispersion,
robotic spectrographs deployed at the 2m Faulkes Telescopes, North and South.
The instruments were designed with supernova classification and
monitoring in mind, with a very large wavelength coverage
($\sim$320 to 1000 nm) and a resolution ($R\sim$300 to 600, depending on
wavelength) well-matched to the broad features, but heterogenous
nature, of these transient events.  The FLOYDS spectrographs are also
excellent for other monitoring programs (e.g.~reverberation mapping of
active galactic nuclei) where the robotic nature of the spectrographs
allow for campaigns not previously possible with classically scheduled
spectrographs.  A more detailed discussion of the FLOYDS instruments,
our design choices, and first science will be presented in a
forthcoming work (Sand et al. in preparation).

\begin{figure}[H]
\begin{center}
\plotone{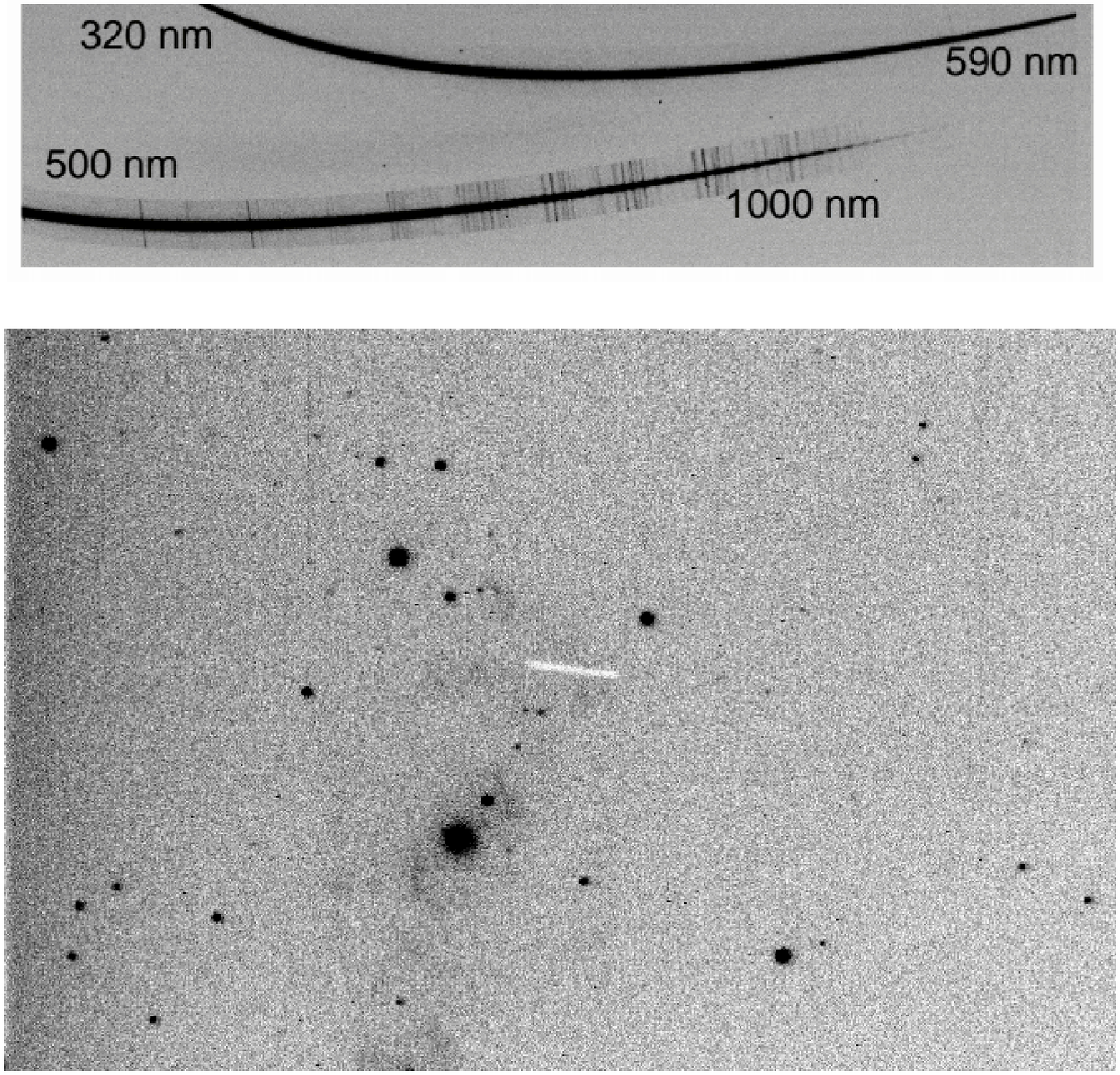}
\caption{Top: A two-dimensional FLOYDS spectrum as seen on the CCD.
  The top trace is second order light, going from 320-590 nm, while
  the bottom trace is $\sim$500-1000nm. Bottom: An image from the
  FLOYDS slit viewing camera, with its 4$\times$6 arcminute field of
  view.  The slit is easily seen in the center of the
  image.  \label{floydsformat}}
\end{center}
\end{figure}

The chosen design uses a low dispersion grating (235 l/mm) and a
cross-dispersed prism in concert to work in first and second order
simultaneously. A folded all-reflecting camera
focuses first- and second-order light
onto the CCD.  This allows a $\sim$320 to 1000+ nm wavelength coverage in
two orders in a single exposure.  
A 30'' slit length allows both orders to fit on
the chip with no order overlap (Figure~\ref{floydsformat}, top).
Four such 30'' length slits are available at each spectrograph,
with widths that bracket the median seeing at each site, and a 6''
slit width for spectrophotometry.  
Since the FLOYDS spectrographs are mounted on the 2m telescopes' Cassegrain
rotator stages, the slit may be oriented along any desired position angle
on the sky.
There are plans for atmospheric dispersion compensators on each FLOYDS,
but these are not yet implemented.
An Andor Newton 940 CCD controller
package, with an e2V 42-10 CCD (13.5 $\mu$m pixels and a 2048$\times$512
format), is used with a broadband ultraviolet-enhanced
coating. 
The Andor Newton CCD package is thermo-electrically cooled to
-70 C, with negligible dark current.
Spectra do fringe above $\sim$700 nm, but this is easily
corrected when a flat field is taken at the same telescope orientation
as a science frame. 

Wavelength calibration is accomplished with a Mercury Argon (HgAr)
lamp, and flat fielding with a combination of a tungsten halogen and
high-powered Xenon lamp (whose light beams are combined with a dichroic).
The calibration unit housing these lamps sits in a cabinet near the
telescope with a fiber connection to FLOYDS, using a Polymicro FBP 
broad-spectrum optical fiber for good transmission across the full 
FLOYDS bandpass.
A deployable arm in the
FLOYDS instrument can direct light from the calibration unit into the
spectrograph, and accompanying optics delivers light with an f/10
beam, mimicking the actual Faulkes telescopes.

For thermal stability and weight reduction, we adopted a space-frame
mechanical structure made of Invar 36, similar to the Levy
Spectrometer at the Automated Planet Finder Telescope
\citep{Radovan10}.  The four slits are interchanged via a rotary
stage; slit position repeatability is better than 0.1 pixels.  FLOYDS
is extremely stable to thermal changes, and has never been refocused
since hardware commissioning, although a stepper motor associated with
the collimator is in place in case the need arises.

Robotic acquisition of spectroscopic targets is the key ingredient for
making robotic spectroscopy possible.  FLOYDS uses a slit-viewing
camera and SBIG STL-6303E CCD camera to image a $\sim$4'$\times$6' field around
the slit both for target acquisition and guiding
(Figure~\ref{floydsformat}, bottom).  Automated target acquisition is
accomplished via one of three modes: 1) a direct world coordinate
system solution of the image with a tailored call to astrometry.net
\citep{Lang10}, followed by centroiding on the object closest to the
target's expected position; 2) placing the brightest point source
within the telescope pointing error circle ($\sim$30'') into the slit
(useful, for instance, for standard stars and bright SNe); and 3)
performing a 'blind offset' from a bright star by a set amount, for
cases in which the target's absolute position is not well-defined.

The FLOYDS data reduction pipeline is a python/pyraf script that
performs standard image detrending (bias subtraction, flat field
correction and defringing), spectral extraction, flux and wavelength
calibration, and spectral combination of the two orders.
In the future, we
intend to integrate these procedures more tightly with the
LCOGT image-reduction pipeline, and we
may implement automated transient classification to shorten the
time lag between data taking and classification announcements.

\subsection{NRES}
\protect\label{NRES}

The Network of Robotic Echelle Spectrographs (NRES) will comprise six
identical high-resolution (R$\simeq$53,000), precise 
(radial velocity repeatability $\lesssim \ 3$ m s$^{-1}$)
optical (380-860 nm) echelle spectrographs, each fiber-fed
simultaneously by two 1m telescopes and a Thorium-Argon (ThAr)
calibration source. Thus, NRES will be a single, globally-distributed
observing facility, composed of six units (one at each of six
Network nodes), using 12 $\times$ 1m telescopes. NRES will roughly
double the RV planet-vetting capacity in the USA, and will achieve
long-term precision of better than 3 m s$^{-1}$ with exposures of less than an
hour for Sun-like stars brighter than V = 12.
Our first spectrograph is scheduled for
deployment in spring of 2014, with the full network operation of all 6
units beginning in fall of 2015.

\subsubsection{Spectrograph Design}

\begin{figure}[H]
 \protect\label{fig:layout}
 \epsscale{0.85} 
 \plotone{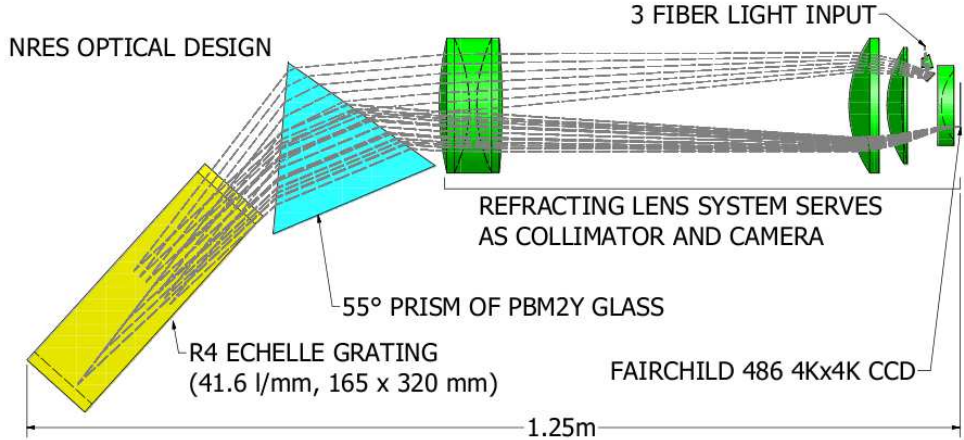}
 \caption{Optical Layout of NRES spectrograph. Starlight is injected via
 a folding mirror at the top right. The field flattener at far right
 doubles as a window to the CCD dewar.}
\end{figure}

The NRES optical design, illustrated in Fig.~\ref{fig:layout}, 
is similar in concept to spectrographs designed for
the Palomar East Arm Echelle \citep{1995PASP..107...62L}, the Lick
Automated Planet Finder \citep{2010SPIE.7735E.153R}, the Carnegie Planet
Finder Spectrograph \citep{2006SPIE.6269E..96C}, and SOPHIE
\citep{2008SPIE.7014E..17P}.
The aim of this design is to achieve very high optical throughput, 
wide wavelength
coverage, and simultaneous fiber input from two telescopes \footnote{
http://lcogt.net/network/instrumentation/nres}.

To maintain wavelength stability, 
we will place the spectrograph in an environmental chamber
that maintains temperature stability of 0.01 C and constant barometric
pressure within 0.2 mb.
These limits will assure Doppler stability (uncalibrated against the
ThAr wavelength standard) at the level of a few tens of ms$^{-1}$,
a factor of 10 better than the unavoidable baseline shifts due to
Earth's rotation.
Only one moving part -- the shutter -- will reside inside this
environmental chamber. By eliminating mechanisms necessary to adjust
optics, we simplify the design, and (more importantly) assure a system
that has great intrinsic stability.
We have tested our shutter design through more than 10$^6$ 
open/close cycles without a failure.

For NRES we will 
use the same CCD controller
as for our standard Sinistro Imagers \citep{2008SPIE.7021E...9T},
but with a different cryostat and re-tuned analog electronics. 
The Fairchild 486 CCD detectors have 4K x 4K format
with 15-$\mu$m square pixels, and are thinned and backside-illuminated,
with broadband antireflection coatings. Their quantum efficiency
peaks at 91\% at 550 nm. The
planned 62.5-$\mu$m input fibers will be over-resolved by a factor of
4.15, oversampling that is very desirable for precise radial velocity
measurement. We expect to achieve a read noise of about 7 e-/pixel at 1
MHz readout rate, requiring 16 s to read the full format.

The analysis pipeline will consist of modules for image calibration,
spectrum extraction, flux and wavelength calibration, radial velocity
determination, stellar classification, quality assurance, and interaction
with the data archive. An important goal in this development is to
minimize human interaction, so that large numbers of spectra can be
analyzed in a very short time, and so that access to the data is easy.

\subsubsection{Prototype}

During the time NRES has been under design, we have also assembled a
scaled-down prototype, which we are using to reduce risk to the NRES
program. The prototype is an R = 25,000 cross-dispersed echelle with the
same basic design as NRES, built mostly from off-the-shelf parts.
We have installed it at LCOGT's 0.83-m telescope at BOS (see \S
\ref{sec:bos}).  The main development goals of the prototype are to
expose unanticipated problems in the general spectrograph design via an
end-to-end test, validate the design of the Acquisition and Guiding Unit
(AGU), which is critical to obtaining high optical throughput, test
hardware and software for the spectrograph's environmental control, and
provide a realistic testbed for the data analysis software. 


We obtained first light with the prototype on the BOS telescope
in October 2012.
Final commissioning and regular operation are awaiting completion of
a major upgrade of the BOS telescope's control systems.
This is due to be finished by May 2013.

\subsection{Extinction Camera}


Each 1m telescope is equipped with an Extinction Camera.  This
wide-angle optical sky imager is designed to provide the
climatic ``context'' of each observation, i.e., data on cloud coverage,
transparency and extinction.
We use Santa Barbara Instrument Group STL-6303E CCD Camera
with a format of 3072$\times$2048 9-$\mu$m pixels. 
A filter wheel holds five filters, as noted in Table 3.
With a Nikon f/2.8 400mm lens, this camera gives an image scale of 4$\arcsec$.64
pix$^{-1}$, and a total field of view spanning 2$^\circ$.6 x 4$^\circ$.0.
The camera is chilled to -20C by a Peltier cooler with waste heat dumped
to the telescope's liquid cooling system.
In dark skies, this system reaches $V=14.9$ in a typical 40s integration.
We mount the Extinction Camera piggy-backed on the 1m telescope primary cell,
coaligned with the telescope optical axis with an accuracy of a few tens of 
arcmin.
The extinction cameras permit rapid comparison of many standard calibrator
star magnitudes with those expected in clear, photometric conditions,
and hence estimates of the current transparency \citep{2013arXiv1301.3926P}.

\section{SOFTWARE}

In order to simplify development and maintenance, the LCOGT software
architecture is highly centralized.
The network topology is a `hub and spoke' configuration
wherein the organization headquarters near Santa Barbara functions as the
hub, radiating control commands to each of the observatory nodes, and
pulling back status information and astronomical data.
Our software architecture reflects this through the notion of
central processing vs processing done at the observatory nodes.  Tasks
such as observation request handling, telescope scheduling,
configuration management, final data reduction, and data archival are handled
at headquarters.

Observatory nodes, telescopes, and instruments are treated as stand-alone
slaves which must robustly and autonomously execute the instructions
they have been given, including operating independently for a period if
network connection to node is lost. Each node must return observation
completion status and updates to technical or weather issues that affect
the evolving global schedule. Node and schedule monitoring tasks are designed
to be comprehensive for the automatic systems, and infrequent for
humans.

As shown by its reliable operation of eight telescopes at four nodes
(including the Back Parking Lot node, in Santa Barbara) for several months,
the software needed to operate the Network is now mature enough to perform
its functions in a workable fashion, under most circumstances.
Nevertheless, we expect to continue vigorous software development for
many years, as
all software systems evolve to provide more functionality,
better monitoring, better and simpler interfacing between components,
fewer failure modes, and better self-correction.

\subsection{Proposal Path}

As with all other aspects of LCOGT operation, the process of accepting,
scheduling, queuing, and executing observation requests will be
highly automated.
The primary point of contact for users of LCOGT will be a Web Portal.
Through the Web Portal, users will submit proposals for evaluation by the
Time Allocation Committee
(TAC), view the status of pending proposals, request observations for
proposals that have been awarded time, monitor pending observation
requests, and access data from observations.

The TAC process will allocate time and assign priorities to each proposal,
based on scientific merit.
Once a proposal has been awarded time, observation requests 
(specifying a particular target, generally along with timing 
or other constraints)
can be added
to that proposal at any time, for execution on the network.  
As
observation requests are added to a proposal, they will be digested by the
Scheduler to be factored into the observing plan, and slated for
execution on particular telescopes throughout the network.

When an observation has been obtained, science data will flow 
through the LCOGT central hub to be calibrated, and then passed on to a Science
Archive (see section 6.8) where users can retrieve their data via various
web interfaces.  Observation status alerts and links to data will be
integrated into the Web Portal for a simple single point of access.

\subsection{Observation Requests and Scheduling}

Perhaps the greatest challenge for the Network design is
the need to schedule requests for disparate kinds of observations -- each
with its own constraints and TAC-assigned priority -- so as to maximize LCOGT's
scientific productivity.
Some aspects of the scheduling operation seem clear.
For instance, to be sure that the telescopes will always be kept busy,
it is desirable to maintain a varied pool of low-priority observing programs
that can be executed in short fragments;  in this way the scheduler will
always have programs available to fill gaps between higher-priority
observations. 
Similarly, sky flats and other calibration data needed by all users should
be scheduled as programs that are owned by LCOGT, and that are automatically
placed into the times when they must occur, at high priority.
Such strategic choices do not, however, speak to the tactical choices that
a scheduler must make to interleave programs with complicated temporal
constraints into a whole that makes best use of the observing resources.
Indeed, a central part of the problem is to define what is meant by
"best use" in this context.
Also, the nature of "best" algorithms and their outcomes will depend
on the mix of observation types desired by users --  a mix that is difficult
to predict and that is likely to evolve in time.
For these reasons, we expect that scheduler development will continue
as a major activity at LCOGT for the foreseeable future,
and we have so far eschewed predictions of the scheduler's efficiency.
In the following paragraphs, we therefore describe the scheduler's
appearance to the user in some detail, but we say little about specific
algorithms used in it (by design, these are modular and
changeable), or about its current performance.

In LCOGT's terminology, the minimum
schedulable unit of the Network is the Block. 
Each Block is bounded in time, and
consists of a number of Molecules, defined as one or more exposures with
a single instrument configuration, at a single telescope
pointing. Multiple Molecules may be grouped within a single Block. 
This structure allows both the user and the Network Scheduler to
operate at a level above raw exposures to define a useful unit of work
in the context of any particular proposal.

Observation requests must include target coordinates and filters, and
may include constraint specifications such as seeing, transparency,
airmass, lunar phase and angle. Requests have a priority assigned by a
TAC. High priority requests, including target of opportunity
observations, can pass quickly though the portal and scheduler to a node
and telescope, possibly interrupting previously scheduled observations.
An interrupted block may resume automatically if the interruption is
short, or may be rescheduled.

Users specify observation
requests in a flexible grammar that allows
arbitrary groupings of individual observations into compound requests
for evaluation by the Network Scheduler.  For example, a user conducting
periodic monitoring of a variable source may specify a series of
observations of a common target at specific times or with approximate
spacings in time, and group these observations together 
as a single compound request.
This says to the Scheduler that all the observations in the compound
request must be obtained in order to satisfy the request.  The Web
Portal provides a graphical environment for composing such requests, but
ultimately requests are transferred in a machine-readable format 
(JavaScript Object Notation),
which provides for programmatic submission as well.


Observation requests may be submitted to the network either manually via
the web portal or programmatically via an Application Program Interface layer.
These submissions
are stored in the Request Database, where they become visible to the
Network Scheduler.  As requests are processed and observations obtained,
their status in the Request Database is updated, and fed back to the
user via the web portal.


The Network Scheduler is responsible for taking observation requests
from the Request Database and distributing them as discrete observations to
be made at specific times on specific telescopes.  Observation requests
may be simple or complex, specific or highly abstract, and may consist
of multiple discrete observing units chained together. In order to
allocate observations to the network, the Scheduler must consider such
factors as program priority, observing constraints, resource
availability, and changing weather conditions.
With these constraints in hand, the Scheduler takes the set of 
incomplete Requests and uses them to
produce a schedule of Blocks for each telescope.

\subsection{Proposals and Observations Network Database (POND)}

Once a schedule has been produced, the corresponding Blocks are stored
in the Proposals and Observations Network Database (POND). This database
acts as the transport layer between the Network Scheduler and the
Network nodes.  Blocks are picked up from the POND by each Site Agent,
and handed to their respective telescopes for execution. 
Since Blocks are the smallest schedulable unit, all Molecules within
a Block must succeed for the Block to be counted as successful.
As the
status (successful or unsuccessful) 
of scheduled Blocks becomes known, they are updated in the POND,
allowing the Scheduler to update the execution of the schedule. The
schedule may be dynamically recalculated as necessary, in
response to Block status information, to accommodate new requests, or to
handle changing network state such as weather conditions or technical
failures.

\subsection{Node Software}

Observations arrive at nodes for execution by a ``pull''
mechanism.  The Site Agent at each node is responsible for checking the
POND and pulling down the node-schedule.  This update happens regularly
as long as network communications between the node and the POND are
good, but in the event of a network outage, the Site Agent maintains a
cache of the latest schedule so it can continue to execute for up to 72
hours without outside connection.

The Site Agent pushes individual observations to a Sequencer agent for
each telescope when conditions are correct for executing the next block
in the schedule.  The Sequencer controls all aspects of the telescope,
instrument, and enclosure necessary to accomplish the observation.  It
manages a large dependency tree to ensure all systems are in proper
states for any given observation before data are collected.  Observations
result in the creation of data artifacts which are analyzed in real
time, and success or failure of the observation is communicated back to
the Site Agent.  Ultimately, the Site Agent pushes observation status
back to the POND so that the Scheduler can perform bookkeeping and
revise the schedule as necessary.

Each component of the system constantly collects telemetry data,
which accumulate in a database at site. These data can
be graphed on the engineering interface to facilitate immediate or
historical diagnoses of events at site in fine detail. These telemetry
data are subsequently resampled into coarser detail and replicated back
to a central telemetry database for performance analysis over longer
time periods.

\subsection{Telescope Control System}

There are no particularly new technologies in the design of our
telescopes, but their control software assumes a very high level of
reliable, autonomous behavior.

The software architecture at each telescope node (controlling the
node, weather and
safety systems, enclosures, telescopes, and instruments) comprises a
Java-based Telescope Control System (jTCS) incorporating a loose collection of
many semi-autonomous agents, developed in the Java Agent DEvelopment
framework (JADE).
Each agent is responsible for a small part of the overall control and
monitoring system, and all participate in a publish and subscribe (PubSub)
system to share data and send each other messages.  This distributed
architecture simplifies the overall control semantics and makes it more
robust against failures:  if any agent dies, the remainder of the
system can continue to operate and attempt recovery of failed
components.

The main agents for each telescope include a) an Astrometric and guiding
agent to configure multiple instruments in the focal plane, based on
Tpoint and the TPK kernel (Terrett 2006); this uses Astrometry.Net for
automatic WCS fitting and to place and guide spectroscopic targets; b)
axis control agents to servo the telescope to the latest target
coordinates; c) agents to monitor International Earth Rotation and
Reference Systems Service (IERS)
bulletins and to configure each
telescope and instrument in the focal plane; 
d) agents to control all enclosure and
telescope systems.  Guiding
corrections can come from any instrument, including
spectrograph slit-viewers, or the science instrument itself.

LCOGT has developed a comprehensive embedded mechanism control system
based on the Blackfin microprocessor family. This system enables internet
control of motors, fans, dry-air systems, mechanisms such as focus,
collimation, filter wheels, mirror covers, and sensors such as
temperature and position probes.  The Blackfin architecture also enables
us to design "smart" power modules to support power cycling and current 
monitoring of each
subsystem, via a JADE agent.  These are key
components of the telescope system's ability to autonomously recover from 
errors and outages.

Each 1m telescope provides support for up to four cooled instrument
electronics crates below each mirror cell, for control of all
instrumentation, fans, sensors and monitoring equipment. The 0.4m
telescopes support similar functionality with fewer control modules, but
with identical servo mechanisms. Our 2m telescopes are 
in the process of off- and on-mount control systems upgrades,
so they too can be integrated into
this PubSub agent control system, and become nodes of our global
network.

\subsection{Configuration Database}
\protect\label{configDB}

For a network as large as LCOGT's, the number of items such as telescopes,
mirrors, cameras, and filters that are deployed in the field 
and that could potentially impact
data quality and provenance runs into the thousands. In order to tackle this
problem of asset management, we have created a database system, called the
ConfigurationDB (ConfigDB) using the Django web framework.

The ConfigDB records details of the sites, enclosures, telescopes, instruments,
cameras, filters, and also several wavelength-dependent quantities 
such as filter transmissions,
CCD QE curves and mirror reflectivities. It also stores start and end dates of
operation, status, and other data relevant for all of these items. Website and
programmatic interfaces allow users to find the canonical
details on the equipment used to take their data, and to perform tasks such as
automatically producing filter transmission plots.

\subsection{Data Pipelines}
\protect\label{datapipelines}

A large network of telescopes such as LCOGT's that will be 
used for a very diverse
set of scientific goals raises unique challenges that are not present in a
single-purpose survey or traditional common-user facility. The large number of
instruments and the volume of data they will generate means that LCOGT, as the
data originator, is in the best position to understand and to reduce the data
optimally. 
On the other hand, the wide variety of scientific programs that will be
running on the network, and their diverse needs for data reduction, renders it
almost impossible to make a generalized pipeline 
optimal for all potential science needs.

Accordingly we have designed the pipeline with the philosophy 
of doing the best we
can for the bulk of potential users, and making the pipeline products 
that are of the
most general use. At the same time, we aim to 
avoid controversial steps
in the data reduction that could be problematic for end users of the 
products, or to attempt to do the end-users' science for them. In addition, the
pipeline emphasizes recording of the processing steps performed, the parameters
used, and the software versions employed. 
These steps are of vital importance
for traceability of the reduced data, and to document its provenance.
The topic of provenance is of increasing importance as astronomical
data sets grow in size and the degree of separation between 
the data producer and
user increases (e.g. \citet{2009arad.workE..27B}).

To provide a pipeline that 
can handle the diverse
instruments of the LCOGT network, 
facilitate adding new instruments, and allow changes
to the type of data products to be made, 
we need a generalized infrastructure that
supplies these capabilities. 
This generic infrastructure is supplied by ORAC-DR \citep{2008AN....329..295C,
1999ASPC..172...11E} which was originally written to support SCUBA 
at the JCMT but
has been extended and generalized (e.g. \citet{2004ASPC..314..460C}) 
to support a
wide variety of instruments and observing types.

The pipeline is entirely data-driven and requires no user input. Processing is
controlled by modular recipes defining the steps
necessary to reduce the data. 
Each recipe is a list of data reduction steps
to perform on each frame or group of frames. These individual steps are known as
primitives, and each primitive performs one 
astronomically-significant step such as 
dark subtraction
or source catalog production. 
As most of the data reduction steps are common across
classes of instruments, 
a small set of primitives is sufficient for the majority
of processing needed in the LCOGT pipeline.

The main recipes currently in use handle the combining of raw bias, 
dark and flat
frames into master calibration frames and the processing of 
regular science frames
to produce the BCD (Basic Calibrated Data) products. 
For these science frames we
perform the following operations:
\begin{itemize}
\item Bad-pixel masking
\item Bias subtraction
\item Dark subtraction
\item Flat field correction
\item Astrometric solution
\item Source catalog production
\item Zeropoint determination
\item Per-object airmass and barycentric time correction computation
\end{itemize}

We perform bad-pixel masking, bias and dark subtraction, 
and flatfield correction 
in the normal manner, using the ORAC-DR calibration infrastructure to select the
nearest (in time) calibration frame that satisfies the constraints of binning,
filter, etc. For astrometric solution, we use \textsc{autoastrom} against
the UCAC3 catalog (for 1.0\,m and 0.4\,m data; \citet{UCAC3}) 
or the Tycho-2 catalog
(for  Context Camera data; \citet{tycho2}). 
Source catalogs are produced using the
SExtractor software \citep{bertin96} to perform object detection, 
source extraction
and aperture photometry. 
In total, we output 49 parameters for each detected source,
consisting of positional information along with its estimated errors, 
and information on the
shape and extent and the measured flux and flux error in 
four fixed and two variable apertures.
Following pipeline processing, the BCD products consisting of 
the reduced images and
source catalogs, PNG bitmap versions of the images, and nightly and
data quality control logs are transferred to the Science Archive 
for ingestion and distribution.

\subsection{Science Archive}
\protect\label{sciarchive}

The Science Archive is expected to play a central role in the LCOGT network, as
this is the primary means for scientists and other users to get access to the
data taken by the network. 
As the public face of the data store, it is intended to be 
accessible and intuitive for a wide variety of potential users.

The Science Archive is being built by 
IPAC\footnote{\url{http://www.ipac.caltech.edu}}
based on a software and hardware architecture that were developed for the Keck
Observatory Archive (KOA) and the NASA Star and Exoplanet Database (NStED)
\citep{2010ASPC..434..119B}.

The archive checks and verifies batches of BCD products (processed images
and their metadata, 
source catalogs, photometric data, and frame bitmaps)
from the pipeline
upon reception and then, if they pass these checks,
ingests them into a relational database management system to allow
fast and efficient querying.
In addition, ancillary data products such as the processing logs and 
master calibration
files contained within a data batch are stored and indexed 
for potential retrieval.

Data from science programs can be read only by authorized users during a
proprietary period which defaults to one year;  data owners may also choose
a shorter period, or to waive the proprietary period.
Data taken for educational purposes are immediately available to all users.

One can access the Science Archive both via a graphical 
web-based interface
that allows sophisticated searching, 
filtering and plotting tasks,
as well as via a programmatic interface. 
This programmatic interface will also be used
to integrate archive queries and functionality into applications developed and
hosted by LCOGT such as those for public education and Citizen Science (see
Section~\ref{education}).
Since the main Archive does not store raw data images, these images are
available to users on request, through LCOGT servers operating at our
Santa Barbara headquarters.

\section{ASSEMBLY AND DEPLOYMENT}

Time-domain observing places unusually stringent demands on
obtaining reliable and consistent behavior from the Network's components.
To meet these demands, we adopted an approach in which LCOGT itself
carried out almost all of the major development tasks.
Thus, most major system components (site development, enclosures, telescopes,
instruments, computing hardware and software) were designed, fabricated,
integrated, tested, and deployed by LCOGT personnel.

\subsection{Prototyping}

We make heavy use of prototypes in a design-fabricate-integrate-test-redesign
cycle to enable full unit and integration testing before major Network
units leave the factory floor.
At an early stage in the design process for each LCOGT major system,
a collaborating team of engineers develop a prototype.
This team unit-tests each subunit to ensure components are appropriate for 
the demands of a fully remote and robotic system, 
to set control sequences and limits, route and protect cabling, 
manage thermal outputs, streamline manufacturing procedures, 
and ensure performance and function. 
When the prototype functions satisfactorily,
it is integrated into the full test observatory system 
which has been configured at the headquarters near Santa Barbara, in a site 
affectionately known as the Back Parking Lot or BPL.
Integration testing places the prototype into the full observatory context 
with current software, enclosure, and other equipment and services. 
The team can then evaluate and improve
the installation and servicing procedures,
minimize thermal footprints, and route cabling.
They then run the  devices through test sequences that replicate normal 
operations and, 
where appropriate, operate the observatory on-sky to ensure 
design specifications are met for alignment, optics, imaging, and performance.
When the prototype is finalized, it is typically upgraded to 
the latest revision and then used as a test device at headquarters to 
help locate and manage
any problems encountered with production devices later on. 
The team updates
detailed plans and bills of materials, and stores them in 
our version-controlled, parts management system from Arena Solutions. 
Purchasing accesses the latest revisions from these systems
and the Logistics and Telops Teams 
maintain a stock of spares of key parts and assemblies, 
both at headquarters and on-site. 
Last, we push any design modifications back into Arena 
and issue change orders as needed for production devices.

\subsection{Assembly}

Once a prototype has been proven in the full observatory, 
the development team reviews the plans and signs off on them. 
The team then has
an adequate number of parts machined, fabricated, or purchased. 
Where necessary, the full anticipated build-out of parts 
(e.g., mirror sets for 15 $\times$ 1-meter telescopes, filter sets, CCDs, etc.) 
is acquired to ensure manufacturing consistency and lower cost. 
Normally, astronomical telescope developments are one-off efforts;
in such cases this elaborate prototype and documentation strategy
would not be cost-effective.
But to produce a dozen or so identical units for a network,
procedures like these are not only essential for correct performance,
they also offer economies of scale that significantly lower overall cost.

We build all observatory components, including the IT support building and 
the enclosures, at LCOGT headquarters in one of four 
assembly bays - enclosures, electrical panels, 0.4m telescopes 
and 1m telescopes. 
We run full unit tests on each major component and integrate the primary 
systems - computing, electrical, and cooling - 
prior to breakdown for shipment.
Each item is then broken down into its major components for shipping to site. 
The assembly crew consists of a project engineer and telescope technicians 
working closely with the design engineers. 
The participation of project engineers ensures that nuances of alignment, 
calibration, assembly, and cabling are communicated and documented. 

\subsection{Deployment}

Site and observatory development follow a standard sequence. 
We first complete
site civil engineering, including concrete and conduit work. 
For the 1-meter telescopes, we then ship the Ash-domes and enclosure walls 
to site and assemble them. 
In the next stage, we install the electrical and computing systems. 
This includes the site IT rack, the telescope and dome control panels, 
the weather station, the air conditioning units and the site air compressor. 
We then ship
the telescopes to site, reassemble, integrate, 
align and collimate them, and test them on sky.
One of the project engineers responsible for final assembly and testing 
at headquarters travels with a team of engineers and telescope technicians 
to site for the installations. 
The 0.4m telescopes, when they are deployed, 
will follow a similar site installation sequence.

Due to the extensive prototyping and testing at headquarters, 
first light on 1m telescopes has typically been 
achieved within a week of installation. 
Robotic control is possible within an even shorter period. 
During the six months following first light, 
we combine engineering evaluation
with internal science use to validate the overall system.

\subsection{Telops}

Telescope operations (Telops) for a global robotic system is 
a critical factor in maximizing the available observing resources and 
in maintaining cost-effective observatory operations. 
LCOGT maintains an experienced crew of Telops experts 
in the UK, California, Hawaii, and Australia.
Moreover, Network nodes are located only at established observatory sites,
where routine maintenance and 
non-critical service or replacement tasks can be completed 
by local staff on a contractual basis. 
We store
a reasonable stock of spare parts, tools, and other necessary gear (such as 
scaffolding) at each site. 
Finally, we take pains to identify performance problems early, through
a comprehensive set of software agents 
that provide status, error reporting, and direct observation and control 
of these remote facilities.

\section{EDUCATION AND OUTREACH}

\protect\label{education}

The aim of the LCOGT education and outreach program is to excite, 
inspire and encourage learners of all ages to pursue science 
investigations and develop their critical thinking skills. 
We are doing this by constructing programs that 
immerse educational users of
our telescopes in fully-formed observational projects that can engage
users at many levels of commitment and sophistication.
A typical project takes participants through observation 
planning and scheduling, image inspection, processing, and analysis,
and publication of the results, either on our 
website or to an external organization, e.g., the Minor Planet Center.
We approach these goals in three ways:  by interacting with individual
participants, with organizations, and by creating and 
distributing web-based resources.

\subsection{Citizen Science}
Citizen Science is often synonymous with 'crowd sourcing', where many people 
repeat a trivial activity a small number of times to produce a more accurate 
result than a small number of people doing the same many times. 
Citizen Science can take this idea further, allowing lay-people to participate 
in scientific investigation without specialized knowledge or experience, 
but training them to think like scientists by having them go beyond mere
acquisition of data, to a meaningful
analysis of the results.

As a pilot to this citizen science approach we created 
Agent Exoplanet\footnote{\tt{ http://lcogt.net/agentexoplanet}},
a browser-based application 
which guides participants as they perform photometry on selected 
exoplanet transit images, 
produce lightcurves, and draw basic conclusions from these data. 
We are currently evaluating this resource and considering which other areas 
of astronomy are adaptable to this approach, to make the most effective use 
of our telescope network with our growing cadre of citizen scientists.

Key to these citizen science projects is the idea that an individual can make 
a contribution to the project, however great or small, which shapes the 
overall outcome. 
In Agent Exoplanet each person can analyze image data, but this analysis is 
then pooled to produce a master lightcurve for each exoplanet transit. 
This fosters the idea that not only is an individual's contribution valuable, 
but that each person's best effort when analyzing the data improves 
the end result for all.

\subsection{Education and Citizen Science Partners}
As part of the LCOGT citizen science program we will support a small number 
of projects such as Agent Exoplanet at any one time, however we will also
encourage external 
groups and organizations to make proposals to our Time Allocation Committee 
for new educational or citizen science projects. 
This will be a parallel route to the science proposal process, 
in which proposals will be judged on their educational and/or 
scientific merits, 
how well they make use of the global nature of our network and how 
they will support their user base. 
Our present plans call for dedicating about 10\% of the observing time
on the 2m telescopes to education projects,
a smaller but as yet undefined fraction of the 1m time,
and 50\% of the time on the 40cm telescopes.
Individuals wanting to use our network will be able to sign up either to
one of the LCOGT programs or to a program supported by external organizations, 
depending on the type of observing experience they desire.

Currently we are working with three organizations, 
our pilot education partners, 
in their use of our expanding telescope network. 
Each of these organizations, 
the  Faulkes Telescope Project (UK), University of Hawaii astronomy outreach 
program (US), and Space to Grow (AU), 
support distinct educational programs using our telescope resources. 
We have worked with these education partners on teacher training programs, 
partnering schemes between astronomers and schools, 
and carried out scientific investigations monitoring asteroids, NEOs, and 
supernovae.

We are focusing effort on making our telescope resources versatile
with intuitive, web-based observation request and simple data analysis
tools. By making these resources available to education organizations
to run their own programs, LCOGT can have a broader and more
sustainable reach in the community than if we were to run all
programs in-house.

\subsection{Online Astronomy Resources}

In addition to creating programs centered around astronomical observing 
and data analysis, an important aspect of our education program is 
to provide web-based tools to help people explore astronomy. 
The following three are a selection of our most popular resources, 
which have been used by school children and teachers, 
by amateur astronomers, and by the general public to enrich their 
experience of astronomy.

\noindent {\bf {Star in a Box}} explores the lifecycle of stars,
a common topic 
in school curricula. 
We developed this interactive Hertzsprung-Russell 
diagram\footnote{\tt{http://lcogt.net/education/starinabox}} as a resource 
for teachers wanting extra support when teaching this topic.

\noindent {\bf {SpaceBook}} is an online astronomy textbook 
written by us. 
It covers most aspects of astronomy with insights into areas 
of expertise provided by our science and technical 
teams\footnote{\tt{http://lcogt.net/spacebook}}.

\noindent {\bf {Virtual Sky}} is a planetarium web application which can 
be embedded into any website. 
We developed this to be easily customizable for viewing location, 
sky projection and many other 
features\footnote{\tt{http://lcogt.net/virtualsky}}.

Full details of our education and citizen science programs can be found 
on our website: {\tt http://lcogt.net/education}.

\section{FIRST SCIENTIFIC RESULTS}

\begin{figure}[!h] 
    \epsscale{0.85}
    \plotone{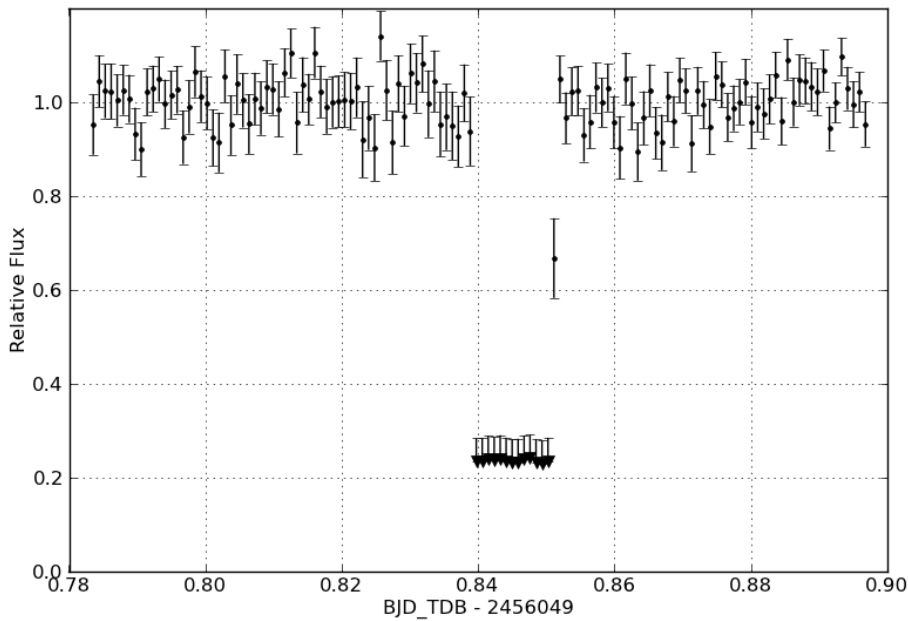}
    \caption{White dwarf light curve taken at the ELP 1.0 m. 
Observations were done with no filter and the exposure time was 60 sec. 
The in-eclipse measurements are upper limits as the object was not detected 
on the CCD in those exposures.}
    \label{WDlc}
\end{figure}

\begin{figure}[!h] 
    \epsscale{0.85}
    \plotone{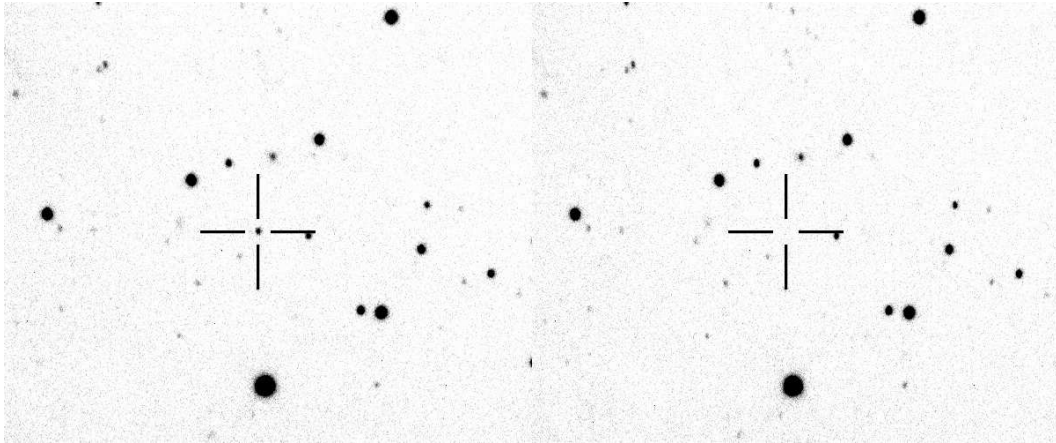}
    \caption{Out-of-eclipse (left) and in-eclipse (right) imaging of 
the white dwarf vicinity, done at the ELP 1.0 m. 
The object was not detected on the CCD in the in-eclipse exposures.}
    \label{field}
\end{figure}

On the night of April 29, 2012, less than one month after first light, 
we used the first McDonald 1m telescope to obtain a light curve of 
a new white-dwarf-containing binary showing deep eclipses. 
We used no filter in order to gain sensitivity, with an exposure time 
of 60 seconds and a total cycle time of 75 seconds. 
The light curve is shown in Fig.~\ref{WDlc}, displayed as 
relative flux vs.~time. 
The object was not detected on the CCD during eclipse, so the in-eclipse 
measurements are marked by the 3-sigma upper limit on the CCD sensitivity. 
The left panel of Fig.~\ref{field} shows an out-of-eclipse image 
centered on the target, while the right panel shows an 
in-eclipse image, where the target cannot be seen.

The duration of the eclipse, combined with multi-band light curves obtained 
at other LCOGT telescopes, indicate the companion is a low-mass star. 
Therefore this system is a possible pre-CV, 
post common envelope eclipsing binary.
A full analysis of all data obtained for this object will be presented 
elsewhere.

These observations were part of a LCOGT photometric follow-up of deeply 
eclipsing WD candidates, with the goal of detecting WDs orbited by 
cold objects, including low-mass stars, brown dwarfs, and planets down to 
sub-Earth size. 
This project is motivated by detections of close-in substellar companions 
to WDs \cite[][]{2006Natur.442..543M, 2006Sci...314.1578L}, 
and many other observations suggesting the existence of planets around WDs 
\cite[e.g.,][]{2009ApJ...694..805F, 2012arXiv1205.3503D}. 
The large amount of LCOGT telescope time and future technical capabilities, 
including the network mode, will be valuable to quickly and efficiently 
confirm and characterize deeply eclipsing WD candidates, 
and to characterize their substellar companion population, in turn 
constraining scenarios for binary system and planetary system evolution 
beyond the main sequence \cite[e.g.,][]{2010MNRAS.408..631N}.

\section{SUMMARY}

In 2012, LCOGT began its transition from designing and
fabricating a global telescope network, to operating it and 
using it for science.
Although the 1m Network will not be complete 
(through Phase B, as defined in section 2) until the end of 2013,
it is clear already that the Phase-B Network, comprising a full ring
of nodes in the southern hemisphere and two nodes (plus additional facilities) 
in the north,
will enable new kinds of time-series
astronomical observations, as intended.
This network of 10 x 1m telescopes is supplemented by new 
low-resolution spectrographs on the Faulkes 2m telescopes,
by a number of 40cm imaging telescopes at sites worldwide,
and (in 2014) by radial-velocity-capable echelle spectrographs on the
1m telescopes.
This paper documents these facilities.

Looking forward, many challenges remain for LCOGT and the Network.
On the level of facilities and instrumentation, we still hope to
complete a northern ring of 1m telescopes to match the southern one
(the Phase-C deployment).
Achieving this will require funding from new sources, perhaps including
raising funds from new partners in the Network.
Using the Network efficiently will demand scheduling and other software
support that goes significantly beyond the expected capabilities of
our first-generation scheduler, proposal-tracking tools, and other
operational facilities.
Thus, we expect to maintain a heavy level of software effort for the
foreseeable future.
Finally, we must organize the scientific collaborations necessary to
apply this powerful observational facility in the most direct possible way
to important scientific problems.
Through the early stages of developing the Network, we were guided in
such matters
by our Scientific Advisory Committee, a rotating group of outstanding 
astronomers from outside LCOGT.
Now that full operations are imminent, we must seek a more direct
connection with our scientific and organizational partners.
We began this process in January 2013, with the first meeting of the
LCOGT Scientific Collaboration, which will govern the use of most of
the Network's observing time.

LCOGT's larger transition -- from an organization that is mostly 
concerned with developing
a facility to one that is mostly concerned with operations and science --
will require significant changes in LCOGT's internal structure, and in its modes
of interaction with the astronomical community.
These changes are already underway; we will describe them elsewhere.

We are grateful to all who have contributed over the last six years to 
the success of the LCOGT Network.
These include those who have served on our governance and advisory
committees:
C. Aerts,
J. Beckers,
T. Bedding,
L. Bildsten,
J. Bloom,
J. Farrell,
L. Hillenbrand,
D. Largay,
T. Marsh,
M. Phillips,
D. Pollacco,
D. Queloz,
P. Roche,
D. Sasselov,
M. Skrutskie,
D. Soderblom,
N. Suntzeff,
J. Tonry,
and V. Trimble.
We also thank all of LCOGT's employees, past and present, whom we have not named elsewhere:
J.R. Aquino,
J.D. Armstrong,
I. Baker,
J. Barton,
O. Barton,
M. Becker,
E. Bildsten,
J. Bowen,
J. Bower-Cooley,
N. Brooks
% ...
and to the Scottish Universities Physics
Alliance (SUPA) for their generous support in helping to install telescopes in
the southern hemisphere.
Funding for the NRES spectrographs was provided through MRI Grant AST-1229720
from the National Science Foundation (NSF).
We are also grateful to Dr. Grant Williams, who refereed this paper, for
his careful reading and helpful comments about the manuscript.


\nopagebreak

\end{document}